\documentclass[twocolumn,aps,pra,superscriptaddress,longbibliography]{revtex4-2}

\usepackage{amsmath,amssymb,amsfonts,bbold}
\usepackage{hyperref}
\usepackage{graphicx}
\usepackage{physics}
\usepackage[dvipsnames]{xcolor}
\usepackage{tikz}
\usetikzlibrary{quantikz}

\newcommand{\cC}{\mathcal{C}}

\newcommand{\cE}{\mathcal{E}}
\newcommand{\cF}{\mathcal{F}}
\newcommand{\cG}{\mathcal{G}}

\newcommand{\cK}{\mathcal{K}}

\newcommand{\cO}{\mathcal{O}}

\newcommand{\cR}{\mathcal{R}}
\newcommand{\cS}{\mathcal{S}}

\newcommand{\cX}{\mathcal{X}}
\newcommand{\cY}{\mathcal{Y}}
\newcommand{\cZ}{\mathcal{Z}}

\newcommand{\upi}{\mathrm{i}}
\newcommand{\upe}{\mathrm{e}}
\newcommand{\id}{\mathbb{1}}

\newcommand{\cir}{\textsc{Cir}}
\newcommand{\vecmu}{\boldsymbol{\mu}}
\newcommand{\veci}{\mathbf{i}}
\newcommand{\veca}{{\boldsymbol{a}}}
\newcommand{\Exp}{\mathbb{E}}
\newcommand{\Var}{\mathbb{V}}
\newcommand{\FSN}{\widehat F_{k,\mathrm{SN}}}

\begin{document}

\title{Simulating general noise nearly as cheaply as Pauli noise}
	
\author{Mark Myers II}
\email{mbmyersii@u.nus.edu}
\affiliation{Centre for Quantum Technologies, National University of Singapore, Singapore}
\author{Mariesa H. Teo}
\thanks{Current Address: University of Chicago, USA}
\affiliation{Centre for Quantum Technologies, National University of Singapore, Singapore}
\author{Rajesh Mishra}
\thanks{Current Address: University of Illinois Urbana-Champaign, USA}
\affiliation{Centre for Quantum Technologies, National University of Singapore, Singapore}
\author{Jing Hao Chai}
\thanks{Current address: Entropica Labs, Singapore}
\affiliation{Centre for Quantum Technologies, National University of Singapore, Singapore}
\author{Hui Khoon Ng}
\email{huikhoon.ng@nus.edu.sg}
\affiliation{Department of Physics, National University of Singapore, Singapore}
\affiliation{Centre for Quantum Technologies, National University of Singapore, Singapore}

\date{\today}

\begin{abstract}
Stabilizer simulation of Clifford quantum circuits---error-correction circuits, Clifford subroutines, etc.---on classical computers has played a central role in our understanding of circuit performance. The stabilizer description, however, restricts the accessible noise one can incorporate into the simulation to Pauli-type noise. More general noise, including coherent errors, may have more severe impact on circuit performance than Pauli noise; yet, such general noise have been difficult to access, much less investigate fully, in numerical simulations. Here, through the use of stratified importance sampling, we show how general noise can be simulated within the stabilizer formalism in reasonable time, with non-unitary noise being nearly as cheap as Pauli noise. Unitary (or coherent) noise can require an order of magnitude more time for the simulation, but nevertheless completes in very reasonable times, a drastic improvement over past approaches that typically fail to converge altogether. Our work thus enables detailed beyond-Pauli understanding of circuit performance in the presence of real device noise, which is rarely Pauli in nature. Among other examples, we present direct simulation results for the performance of the popular rotated planar surface codes under circuit-level general noise, previously available only in limited situations and/or through mappings to efficiently simulatable physical models. 
\end{abstract}
	
\maketitle

\section{Introduction}
Simulation of quantum phenomena on classical computers has been essential for furthering our understanding of the quantum realm. Because of the underlying exponentially large state space, classical simulation is tractable only for small quantum systems, unless one can rely on special structural or symmetry properties of the problem (e.g., limited entanglement, restricted operations, etc.) to simplify the representation.
One such example is the stabilizer formalism \cite{gottesman1996,gottesman1997}, which permits efficient classical simulation of large quantum systems, including ones with thousands of qubits---two-dimensional quantum systems. It requires, however, a restriction in the allowed states to just ($n$-qubit) stabilizer states, each a joint eigenstate of $n$ independent and commuting $n$-qubit Pauli operators. In a stabilizer simulation, one begins with a stabilizer state, and the allowed subsequent operations are ones that keep the quantum state within the set of stabilizer states. Such operations are termed stabilizer or Clifford operations, and the sequence of operations is a stabilizer or Clifford circuit. Stabilizer circuits can be efficiently classically simulated, as encapsulated in the Gottesman-Knill theorem \cite{gottesman1997,gottesman1998}.

Stabilizer states are a restricted set, but they include highly entangled states that manifest nonclassical behavior. Stabilizer circuits, while not universal for quantum computing, are used in crucial quantum subroutines like those for error correction with stabilizer codes, and general quantum circuits often comprise large tracts of stabilizer circuits interleaved with a small fraction of nonstabilizer operations. Circuit compilation protocols routinely minimize the nonstabilizer-count of quantum circuits, as nonstabilizer operations can be harder to implement, especially when coupled with error correction.

Ideal stabilizer circuits can be efficiently simulated classically through the stabilizer formalism; noisy or imperfect stabilizer circuits, on the other hand, can be efficiently simulated only if the noise is also a stabilizer operation so that the system is not taken out of the set of stabilizer states by the noise. In fact, most noisy stabilizer simulations look only at a particular class of stabilizer noise: Pauli noise, with $X$, $Y$, or $Z$ single-qubit Pauli errors occurring with specified probabilities, with generalization to more (usually two) qubits. The vast majority of numerical studies of noisy stabilizer circuits are limited to Pauli noise, including the numerical evidence for the favorable performance of popular codes like surface codes and other quantum-LDPC codes explored as routes to reliable quantum computers.

Yet, real devices do not just suffer from Pauli, or even more general stabilizer, noise. Amplitude-damping noise, arising from decay processes, and coherent (or unitary) errors from over- or under-rotation, are two other commonly studied elementary noise models but which take one outside the set of stabilizer states. Real devices see noise that is a combination of these elementary and more general noise processes; modeling of real devices must include such general noise. Notably, past works showed genuine qualitative differences between quantum tasks carried out in the presence of Pauli and coherent noise. Ref.~\cite{wallman2014} pointed out that the often-used randomized benchmarking protocol \cite{emerson2005,knill2008} for assessing quantum gate quality requires many more iterations for accurate estimates of the average infidelity in the presence of coherent errors compared to unital but nonunitary (including Pauli) noise. Refs.~\cite{sanders2015,kueng2016} showed that gate quality, when measured via the diamond distance relevant for fault-tolerant quantum computing, scales as the average infidelity for depolarizing (i.e., symmetric Pauli) noise, but as the square-root of the infidelity for coherent noise; this difference in scaling persists even if one compares with worst-case, rather than average, infidelity \cite{wallman2016}. Coherent noise also resists description as stochastic noise employed in standard proofs for fault-tolerant quantum computing \cite{kitaev1997,aharonov1997,knill1998,aliferis2006}; accommodating coherent noise requires an amplitude---not probabilistic---treatment, which gives an effective squaring of the provable noise threshold below which fault-tolerant quantum computing works \cite{terhal2005,aliferis2006,aharonov2006,novais2007,novais2008,ng2009}. The common belief, however, is that a numerical estimate for the threshold for general noise will give numbers more realistic than the overly stringent theoretical value limited by proof techniques. 

There is thus a clear need to study quantum protocols in the presence of noise beyond simple Pauli noise. This is especially pertinent when assessing the performance of quantum error-correcting codes on which lies the current hope of large-scale accurate quantum computers. Already in 2014, Ref.~\cite{tomita2014} investigated the 17-qubit distance-3 planar surface code for amplitude- and phase-damping noise, making use of full quantum simulation to tackle the beyond-Pauli situation. This severely limited the code size that could be studied. Refs.~\cite{gutierrez2015,gutierrez2016} tried to circumvent this limit by making use of Clifford approximations of non-stabilizer channels, which, if successful, would render general noise accessible via stabilizer simulations, but found the approximations poor at predicting code performance. The tensor-network approach of Ref.~\cite{darmawan2017} saw some success, and allowed simulation of planar surface codes up to distance 11 under code-capacity noise, but the method worked well only for amplitude-damping noise while remaining inaccurate for the unitary $Z$-rotation noise studied. 

A different tack from direct simulation was to map surface codes to physical models that admit efficient simulation. Ref.~\cite{suzuki2017} gave one of the earliest studies in this direction, mapping the repetition code---viewable as a 1D version of the surface code---to a fermionic model and then studying the code under coherent errors through simulating that model. Ref.~\cite{bravyi2018} used an earlier mapping \cite{wen03,kitaev06} of the surface code to Majorana fermions that allowed for simulation with coherent noise up to an impressive distance 37, though only under code-capacity noise and limited to $Z$ rotations only; this was extended to include readout errors in Ref.~\cite{marton2023}. In the last few years, B{\'e}ri and co-workers also used various statistical mechanics mappings to study surface codes in the presence of coherent errors from a phase-transition perspective \cite{venn2020,venn2023,behrends2024,behrends2025}. These studies assumed ideal error correction operations, as did an earlier general-codes work~\cite{huang2019}, to allow for derivation of an effective logical noise channel after error correction to simplify matters; see also the generalization to \emph{qudit} surface codes \cite{ma2023}.

These many efforts, with varying degrees of success, point back at the need for a general approach to simulating many-qubit quantum circuits with beyond-stabilizer noise. The efficient numerical route through mappings to physical models worked well for the surface code because of its deep origins in physical models; one cannot expect such nice mappings to exist for generic codes, or general circuits. The numerics must also be able to capture general noise, not just specific noise like $Z$-rotation coherent error, and for a large class, if not all, of useful circuits.

Bennink et al.~in Ref.~\cite{bennink2017} took a step in this direction. They showed that an arbitrary noise process can be decomposed as a linear combination of stabilizer channels. Using this, they built a Monte Carlo approach based on stabilizer formalism to compute expectation values of observables for stabilizer circuits, including the fidelity measure used in error correction studies. That the decomposition is a linear---not convex---combination calls for quasiprobabilistic sampling to deal with stabilizer channels with negative weights. Therein lies the sign problem well known from simulation of ferminionic systems, and the exponential inefficiency of any such method that steps into the quantum realm beyond the stabilizer domain (see, for example, the exponential inefficiency of simulating the non-Clifford $T$-gate in Ref.~\cite{bravyi2016}).

Nevertheless, for noise with small ``negativity", which measures the size of the negative weights, the method of Bennink et al.~worked well enough for them to simulate the 7-qubit Steane code under amplitude-damping noise. It appears to not work so well, however, for coherent errors, with the only such example given in the paper to be of a single qubit undergoing $Z$ rotation. A more recent work \cite{hakkaku2021} that sought to examine coherent errors using the same approach admixed in bit-flip noise, rather than looking at pure $X$-rotation errors. In contrast, Ref.~\cite{leblond2025} (with R.~Bennink as the last author) used a different decomposition to explore the situation where coherent errors are introduced only at a restricted number of circuit locations, though the precision of the simulations remain limited. Our own attempts at using the Bennink et al.~method to simulate even the smallest surface code in the presence of circuit-level coherent errors simply never converged to even generous-sized error bars after waiting for a long time. The method did work reasonably for nonunitary, but non-stabilizer, noise though, already for the $d=5$ surface code, the time taken was significant; it was clear one cannot scale much beyond the tens-of-qubits size.

The issue here is one of too-large variance, and for that, we can turn to variance-reduction techniques from the vast Monte Carlo sampling literature. In this work, we show how to use stratified importance sampling (see, for example, Refs.~\cite{Press2007,Ross2023}), together with the stabilizer decomposition approach of Ref.~\cite{bennink2017}, to simulate general noise---including coherent errors---within the stabilizer formalism nearly as cheaply as Pauli noise. Our main example is the well-studied surface code, given its high relevance in quantum computing implementations today, and for comparison with existing work. There, we show how stratified sampling can estimate the logical infidelity for the standard depolarizing noise, as well as for a variety of unitary and nonunitary noise. As an example of the speed of our approach, for the case of the distance-7 surface code involving 97 qubits (the subject of the recent experimental demonstration by the Google Quantum AI team \cite{google2025}), estimating the logical infidelity for a single physical infidelity value took between 2 (depolarizing noise) and 13 (a unitary or coherent noise) seconds; the nonunitary noise all took $<5$ seconds (see Tab.~\ref{tab:d7Timing} below). We show similar relative timing costs between unitary and nonunitary noise for up to distance-15 surface codes involving 449 qubits. 

The paper is organized as follows. We begin with the problem setup in Sec.~\ref{sec:Setup}. Section \ref{sec:StdIS} describes the earlier approach of Ref.~\cite{bennink2017} in the language of standard importance sampling, and then we present our stratified importance sampling approach in Sec.~\ref{sec:StrIS}. We then demonstrate, in Sec.~\ref{sec:Examples}, how to simulate general noise nearly as cheaply as Pauli noise using stratified sampling for the Steane code and rotated planar surface code error-correction protocols. At the end, to show general applicability of our technique beyond error correction tasks, we also discuss a use case in a recent Clifford error reduction protocol. We conclude in Sec.~\ref{sec:Conc}, and the appendices expand on the technical details described in the main text.

\section{Problem setup}\label{sec:Setup}

We begin with a few definitions. The single-qubit Pauli operators are $X\equiv \ketbra{0}{1}+\ketbra{1}{0}$, $Y\equiv -\upi\ketbra{0}{1}+\upi\ketbra{1}{0}$, and $Z\equiv \ketbra{0}{0}-\ketbra{1}{1}$, where $\ket{0}$ and $\ket {1}$ are the $\pm 1$ eigenstates of $Z$, forming the computational basis. The 1-qubit Pauli group $\cG_1$ is the set $\{s\sigma\}$ with $s\in\{\pm 1,\pm i\}$ and $\sigma\in\{\id, X,Y,Z\}$, and $\id\equiv \ketbra{0}{0}+\ketbra{1}{1}$ is the identity. The $n$-qubit Pauli operators and Pauli group $\cG_n$ elements are tensor products of the single-qubit versions. An $n$-qubit stabilizer state $\ket{\psi}$ is the unique simultaneous $+1$ eigenstate of $n$ independent commuting elements of $\cG_n$. Any $n$-qubit operator can be written as a linear combination of $n$-qubit stabilizer states $\ketbra{\psi}{\psi}$; the minimal number of such states in the linear combination is its stabilizer rank. 

A stabilizer channel $\cS$ is a completely positive (CP) and trace-preserving (TP) operation that maps stabilizer states to stabilizer states. The set of stabilizer channels comprises the Clifford gates, $U_i(\cdot)U_i^\dagger$ with $U_i$ a Clifford unitary operator, and the Pauli-reset channels, each describing the measurement of a Pauli observable, followed by a reset (can be omitted if the circuit terminates there for the measured qubits) to one of the eigenstates of that observable. A stabilizer circuit comprises a sequence of stabilizer channels on $n$ qubits. In the typical situation, the stabilizer circuit has only Clifford gates and perhaps Pauli measurements at the end, but we can have more general stabilizer channels (e.g., classically controlled stabilizer operations). Stabilizer noise refers to noise describable by a stabilizer channel.

Given a stabilizer circuit $\cir$ on $n$ qubits in the initial state $\rho$, we are interested in computing the expectation value of an $n$-qubit observable $\cO$ on the output of $\cir$. In the presence of noise, every operation in $\cir$, including waiting steps, can suffer errors. We denote the noisy circuit as $\widetilde\cir$, and we want the expectation value (assumed to be a real number),
\begin{equation}
F(O,\rho; \widetilde\cir)\equiv \mathrm{Tr}\{O\widetilde\cir(\rho)\}.
\end{equation}
We can write $\widetilde\cir$ as a spacetime sequence of noisy operations (channels),
\begin{equation}
\widetilde \cir=\widetilde\cG^{(A)}\widetilde\cG^{(A-1)}\ldots \widetilde\cG^{(3)}\widetilde\cG^{(1)},
\end{equation}
with $\widetilde\cG^{(a)}\equiv \cE^{(a)}\cG^{(a)}$ for $\cG^{(a)}$ a gate (noise acts after), and $\widetilde\cG^{(a)}\equiv \cG^{(a)}\cE^{(a)}$ for $\cG^{(a)}$ a measurement (noise acts before). Here, $\cE^{(a)}$ is the CPTP noise that acts at the circuit spacetime location $a(=1,2,\ldots, A)$ corresponding to the ideal stabilizer operation $\cG^{(a)}$, and there are $A$ circuit locations in $\cir$. Here, we have assumed it possible to write the noisy $\widetilde \cG^{(a)}$ as the action of a noise channel $\cE^{(a)}$ followed/preceded by the ideal operation; such an $\cE^{(a)}\equiv  \widetilde\cG^{(a)}{\cG^{(a)}}^\dagger$ always exists for unitary $\cG^{(a)}$, but it is a nontrivial assumption for nonunitary $\cG^{(a)}$s (e.g., measurements). Nevertheless, such a noise model is commonly assumed in studies of noisy quantum circuits and applies in many practical situations.

Assuming $\rho$ and $O$ are of $\textrm{poly}(n)$ stabilizer rank, and if we only have stabilizer noise, i.e., all $\cE^{(a)}$s are stabilizer channels, the stabilizer formalism enables efficient---$\textrm{poly}(n)D$---computation of $F$, where $D$ is the depth of $\cir$. We are, however, interested in computing $F$ for general, not necessarily stabilizer, noise. Ref.~\cite{bennink2017} pointed out how to do this using Monte Carlo simulation. Below, we describe their prescription (Sec.~\ref{sec:StdIS}), which amounts to standard importance sampling, and explain how one can run into problems for some noise channels. We then propose our stratified importance sampling approach (Sec.~\ref{sec:StrIS}) that exploits the weak nature of the noise for feasible simulation time for general noise.

\section{Standard importance sampling}\label{sec:StdIS}

To phrase the estimation of $F$ as a sampling problem, we begin with a useful decomposition of CPTP channels: Ref.~\cite{bennink2017} showed that any $n$-qubit CPTP $\cE$ can be written as a linear combination of $n$-qubit stabilizer channels,
\begin{equation}\label{eq:StabDecomp}
\cE = \sum_{\mu=0}^I q_\mu\cS_\mu,
\end{equation}
where $\cS_\mu$ is a stabilizer channel, and $q_\mu\neq 0$s are real scalars with $\sum_{\mu=0}^I q_\mu = 1$ for TP $\cE$. The decomposition is not unique. $\eta\equiv \sum_{\mu, q_\mu<0}|q_\mu|$ is the ``negativity" \cite{bennink2017} of the stabilizer decomposition, and $\sum_{\mu=0}^I |q_\mu|=1+2\eta$.

Using this, we write every noise channel in $\widetilde\cir$ as a stabilizer decomposition: $\cE^{(a)}=\sum_{\mu_a=0}^{I^{(a)}}q^{(a)}_{\mu_a}\cS_{\mu_a}$. In addition, $\rho=\sum_{\mu_0=0}^{I^{(0)}} q^{(0)}_{\mu_0} \varrho_{\mu_0}$ and $O=\sum_{\mu_{A+1}=0}^{I^{(A+1)}}q^{(A+1)}_{\mu_{A+1}}\varrho_{\mu_{A+1}}$, where $\varrho_i$'s are $n$-qubit stabilizer states (density operators), and $\rho$ and $O$ are regarded as locations $0$ and $A+1$, respectively. Then, the expectation value of interest can be written as \footnote{Note that, for simplicity, Eq.~\eqref{eq:F} is written as if all circuit operations are gates, so that errors follow the ideal operation; the order should be reversed whenever the circuit operation is a measurement.}
\begin{align}
\label{eq:F}
F &= \sum_{\mu_0,\mu_1,\ldots,\mu_{A+1}}  q^{(A+1)}_{\mu_{A+1}}q^{(A)}_{\mu_{A}}\ldots q^{(0)}_{\mu_0}\\
& \times \mathrm{Tr}\{\varrho_{\mu_{A+1}}\cG^{(A)}\cS_{\mu_{A}}\ldots \cG^{(1)}\cS_{\mu_1}(\varrho_{\mu_0})\}\equiv \sum_{\vecmu}q({\vecmu})f({\vecmu}),\nonumber
\end{align}
where $\vecmu\equiv (\mu_0,\ldots, \mu_{A+1})$, $q({\vecmu})\equiv q^{(A+1)}_{\mu_{A+1}}\ldots q^{(0)}_{\mu_0}$, and $f({\vecmu})\equiv \mathrm{Tr}\{\varrho_{\mu_{A+1}}\cG^{(A)}\cS_{\mu_{A}}\ldots \cG^{(1)}\cS_{\mu_1}(\varrho_{\mu_0})\}$. We refer to a sequence $\{\varrho_{\mu_{A+1}}, S_{\mu_{A}}, \ldots, S_{\mu_1}, \varrho_{\mu_0}\}$ as a ``configuration", specifying the choice of stabilizer states or channels at each location. Since each configuration contains only stabilizer states and channels, and the ideal $\cG^{a}$s are stabilizer operations, we can compute $f$ using standard stabilizer formalism (see, for example, \cite{gottesman1998, aaronson2004, nielsen2010,gidney2021}).

Ref.~\cite{bennink2017} noted that $F$ can be estimated using Monte Carlo simulation: We sample different configurations, compute $f$ for each configuration with the stabilizer formalism, and then sum over the configurations with the appropriate coefficients to obtain an estimate of $F$. However, since the $q_\mu$s can be positive or negative, a proxy proposal distribution has to be employed for the Monte Carlo sampling. Ref.~\cite{bennink2017} proposed a particular route to this, which we describe below.

The Monte Carlo approach of Ref.~\cite{bennink2017} can be viewed as standard importance sampling (see, for example, \cite{Press2007, Ross2023}). Importance sampling is a general technique to estimate quantities built from an inaccessible distribution by sampling from an easier-to-sample proxy distribution. In our current setting, $F$ is built from $\{q_{\mu_a}^{(a)}\}$s, which do not form probability distributions. Instead, we rely on proxy distributions $\{p^{(a)}_{\mu_a}\}$s, with $\sum_{\mu_a}^{I^{(a)}} p^{(a)}_{\mu_a}=1$ and $p^{(a)}_{\mu_a}> 0$; $p^{(a)}_{\mu_a}$ can otherwise be freely chosen. By drawing $M$ independent samples according to $\{p^{(a)}_{\mu_a}\}$s, we have
\begin{equation}\label{eq:FM}
\widehat{F}_M = \frac{1}{M}\sum_{m=1}^M w({\vecmu^m})f({\vecmu^m}),
\end{equation}
as the standard importance sampling estimator for the expectation value $F$, with each $\vecmu^m\equiv(\mu_0^m,\mu^m_1,\mu^m_2,\ldots,\mu^m_{A},\mu_{A+1}^m)$ specifying a randomly sampled configuration. The $w({\vecmu^m})$s in Eq.~\eqref{eq:FM} are weights to compensate for sampling with the proxy $p$s instead of the $q$s:
\begin{equation}\label{eq:w}
w({\vecmu^m})\equiv \frac{q(\vecmu^m)}{p(\vecmu^m)}= \prod_{a=0}^{A+1}\frac{q^{(a)}_{\mu_a^m}}{p^{(a)}_{\mu_a^m}}.
\end{equation}
The estimator $\widehat F_M$ is unbiased, $\Exp[\widehat F_M]=\Exp[\widehat F_1]=\sum_\veci p(\vecmu)\frac{q(\vecmu)}{p(\vecmu)}f(\vecmu)=F$, and has variance $\Var[\widehat F_M]\leq \frac{1}{M}\Exp[w(\vecmu)^2]$ (see App.~\ref{app:SamplingIS}). 
Thus, to achieve error $\sqrt{\Var(\widehat F_M})$, $M= \frac{1}{\epsilon^2}\Exp[w(\vecmu)^2]$ samples suffice.

$\Exp[w(\vecmu)^2]$ is determined by the proxy $p$. Ref.~\cite{bennink2017} prescribes choosing $\{p_{\mu_a}^{(a)}\}$ to minimize $\Exp[w(\vecmu)^2]=\prod_{a=0}^{A+1}\Exp[(w_{\mu_a}^{(a)})^2]$. The constrained minimization problem is easily solved to give $\{p_{\mu_a}^{(a)}= |q_{\mu_a}^{(a)}|/\alpha^{(a)}\}$, with $\alpha^{(a)}\equiv 1+2\eta^{(a)}$, where $\eta^{(a)}$ is the negativity for location $a$. The weights are then $w_{\mu_a}^{(a)}\equiv q^{(a)}_{\mu_a}/p^{(a)}_{\mu_a}=\mathrm{sgn}(q_{\mu_a}^{(a)})\alpha^{(a)}$. If $\eta^{(a)}\sim \eta~ \forall a=1,\ldots, A$, similar in size for all circuit locations, the variance for this proxy choice scales as
\begin{equation}\label{eq:VarBennink}
\Var[\widehat F_M]\lesssim \frac{1}{M} (1+2\eta^{(0)})(1+2\eta^{(A+1)}) (1+2\eta)^{2A},
\end{equation}
exponential in the size $A$ of $\cir$. For large circuits, this entails requiring a large sample for an accurate estimate of $F$, with the problem becoming more severe the larger the negativity $\eta$. That accurate estimates become expensive should come as no surprise---we do not expect to efficiently simulate an arbitrary quantum setting on a classical computer. 

Nevertheless, for small enough circuits and small enough negativity, as shown in Ref.~\cite{bennink2017}, and later in Ref.~\cite{hakkaku2021}, one can carry out the above prescription to estimate $F$. Our own attempts at this worked well for simulations of small-sized problems in the presence of nonunitary noise of strengths relevant for real devices. For similar-strength unitary noise, however, the estimation error remained large even after an extremely large number of samples. The difference in simulatability between unitary and nonunitary channels may be partly explained by the fact that unitary noise has negativity and noise strength that decrease more slowly with channel infidelity (see Fig.~\ref{fig:UnitVsNonunit} in App.~\ref{app:UnitVsNonunit}). We also observed that the negativity of the noise was not the sole predictor of the number of samples needed---unitary noise seems to require many more samples than nonunitary noise with the same negativity. While the upper bound on the variance above [Eq.~\eqref{eq:VarBennink}] depends only on $\eta$, the number of samples needed in an actual simulation also depends on the variance of $f$ across different configurations, something neglected in the bound.

Below, we explain how to make use of the weak nature of the noise---ignored in the prescription so far---to reduce the impact of the variance of $f$ on the estimation error.

\section{Stratified importance sampling for weak noise}\label{sec:StrIS}

In our setting of interest, the $\cE^{(a)}$s describe noise in circuit components, and are weak---close to the identity---in any device useful for quantum computation. Thus, it is reasonable to assume (we see below that there is in fact no loss of generality here) that the stabilizer decomposition for $\cE^{(a)}$ takes the form
\begin{align}
\label{eq:WeakNoise}\cE^{(a)}\!=\!{\left[1\!-\gamma^{(a)}\right]}\id+\gamma^{(a)} \cF^{(a)},
\quad\cF^{(a)}\!\equiv\!\! \sum_{i=1}^{I^{(a)}}r_{i}^{(a)}\cS_{i},
\end{align}
with $\gamma^{(a)}\in[0,1]$ (typically $\ll 1$) the noise strength at location $a$, $\id\equiv \cS_0$ is the identity channel, and $\cF^{(a)}$ is a TP (though not necessarily CP) channel composed from the nonidentity stabilizer channels $\cS_{i\neq0}$. In the earlier $q_\mu$ notation, $q_0^{(a)}=1-\gamma^{(a)}$, and $q_{i>0}^{(a)}=\gamma^{(a)} r_{i}^{(a)}$. We use the convention that the indices $\mu$s go from $0$ to $I^{(a)}$, while $i$s go from $1$ to $I^{(a)}$. Note that $\sum_{i} r_{i}^{(a)}=1$ so $\cF^{(a)}$ is TP. 

The form of Eq.~\eqref{eq:WeakNoise} motivates the following noise description of the noise: No fault (i.e., $\id)$ occurs at location $a$ with probability $1-\gamma^{(a)}$, while fault $\cF^{(a)}$ occurs with probability $\gamma^{(a)}$; when fault $\cF^{(a)}$ occurs, different errors $\cS_i$s occur with weights $r_i^{(a)}$s. This is reminiscent of how we describe probabilistic noise, but note that $\cF^{(a)}$ need not be a physical channel, the $r_i^{(a)}$s are not always positive, and our current description applies to \emph{all} CPTP noise, probabilistic in the usual sense or otherwise. Note that the form of $\cE$ in Eq.~\eqref{eq:WeakNoise} in fact applies to all channels; see footnote \footnote{If the prefactor of $\id$ in the stabilizer decomposition of $\cE$ exceeds 1 (which, in principle, is possible), but not by much---necessary for weak noise and small enough negativity for plausible simulatability---one can put some of that $\id$ into $\cF$, and define an appropriately small but positive $\gamma$ so that Eq.~\eqref{eq:WeakNoise} holds. We note, however, that in all our numerical investigations of single-qubit noise channels, we never encountered a $>$1 prefactor for $\id$ in the decompositions, at least not for the noise strengths relevant for our examples. This may be a consequence of the r igid conditions on CPTP single-qubit channels (e.g., see Ref.~\cite{king2001}), but we leave that as a conjecture.}.

Let us assume that all circuit locations have the same noise strength $\gamma^{(a)}=\gamma$ for $a=1,2,\ldots, A$, though they can have different $\cF^{(a)}$s. The case of inhomogeneous noise strengths can be easily accommodated (see Apps.~\ref{app:pk} and \ref{app:fault_location_selection}, and Ref.~\cite{myers2025}), but we restrict to this single-parameter $\gamma$ situation here for simplicity. 
Now, in many situations, we discuss the performance of a noisy circuit in terms of the number of faulty locations. For example, in error-correction settings, we consider configurations as benign when no more than $t$ faults occur in the circuit, with $t$ as the maximal number of errors tolerated by the code. We may also be concerned with the accuracy of a calculation to order $\gamma^\ell$, so that we need not consider more than $\ell$ insertions of $\cF$s across the circuit. It thus seems useful to re-organize---or ``stratify"---the sum over all configurations $\vecmu$ in $F$ [Eq.~\eqref{eq:F}] according to the number of faults $k$. 

Let $\cC_k$ denote the set of configurations where faults ($\cF^{(a)}$s) are inserted at exactly $k$ circuit locations; $\id$ is inserted at the remaining $A-k$ circuit locations, while all possible choices of $\varrho_i$s (present in the decompositions of $\rho$ and $O$) can be inserted at $a=0$ and $A+1$. We denote the configuration by the index $\veci_\veca\equiv (i_0,i_{a_1},i_{a_2},\ldots, i_{a_k},i_{A+1})\in\cC_k$, where $\veca\equiv (a_1,a_2,\ldots, a_k)$ specifies the $k$ faulty locations and $|\veca|=k$. We refer to $\cC_k$ as a ``stratum" and observe that $\bigcup_{k=0}^A\cC_k$ gives all configurations. 
The sum in $F$ can then be re-organized as
\begin{equation}
F=\sum_{k=0}^AP_\gamma(k)F_k,\quad F_k\equiv \binom{A}{k}^{\!-1}\!\!\!\sum_{\veci_\veca\in\cC_k}r(\veci_\veca)f(\veci_\veca),
\end{equation}
where $P_\gamma(k)\equiv$ the probability of exactly $k$ faulty locations $= \binom{A}{k}(1-\gamma)^{A-k}\gamma^k$ in our case of homogeneous noise strength. $F_k$ is the value of $F$ restricted to stratum $\cC_k$ (i.e., conditioned on there being exactly $k$ faulty locations in the circuit), and $r(\veci_\veca)\equiv q_{i_0}q_{i_{A+1}} \prod_{a\in\veca}r_{i_a}^{(a)}$. Furthermore, as mentioned earlier, not all $k$ values may be relevant (e.g., $k$ up to some $k_{\max}$ like $t$ or $\ell$), or we may have prior knowledge of $F_k$ for some values of $k$. To accommodate this, we let $\cK$ denote the set of relevant values of $k$ to be sampled over, and write
\begin{equation}
F=(1-P_\cK)F_{\notin\cK}+\sum_{k\in\cK}P_\gamma(k)F_k,
\end{equation}
where $P_\cK\equiv \sum_{k\in\cK}P_\gamma(k)$, and $F_{\notin\cK}$ is a known value that accounts for $F_k$s when $k\notin\cK$. $\cK$ can be the full range if there is no prior knowledge and all $k$s are relevant.

This form of $F$ suggests a sampling strategy that stratifies according to $k$: For each $k\in\cK$, estimate $F_k$ using importance sampling with $M_k$ independent samples from $\cC_k$ and the standard estimator [Eq.~\eqref{eq:FM}], 
\begin{equation}\label{eq:FkMk}
(\widehat F_k)_{M_k}=\frac{1}{M_k}\sum_m w(\veci_\veca^m)f(\veci_\veca^m),\quad \veci_\veca^m\in\cC_k.
\end{equation}
The full $F$ is then estimated as
\begin{equation}
(\widehat F_\textrm{str})_M=(1-P_\cK)F_{\notin\cK}+\sum_{k\in\cK}P_\gamma(k)(\widehat F_k)_{M_k},
\end{equation}
with $M\equiv \sum_{k\in\cK}M_k$. We refer to $\widehat F_k$ as the ``stratum estimator", and $\widehat F_\textrm{str}$ as the ``stratified estimator". It is easy to see that the $\widehat F_k$s, and hence $\widehat F_\textrm{str}$, are unbiased estimators. The conditional distribution $p(\veci_\veca|k)$ chosen as proxy for $r(\veci_\veca)$ with $\veci_\veca\in\cC_k$ in the importance sampling, as well as the $M_k$ values, can be optimized to minimize the variance of $\widehat F_\textrm{str}$. Appendix \ref{app:SamplingSIS} gives the details, following a similar logic as in standard importance sampling.

Splitting up the sampling according to $k$ falls under the well-known Monte Carlo technique of stratified sampling   (used also very recently in Ref.~\cite{heussen2024}, under the tagline of subset sampling, to study quantum error correction protocols). Stratified sampling is a common variance-reduction approach, guaranteed to yield a smaller variance than the unstratified version, and especially effective when $F_k$ depends strongly on the value of $k$ (though it generally does not reduce the formal complexity of the sampling problem); see App.~\ref{app:SamplingSIS}. In error-correction settings, where one chooses $F$ to be the fidelity of the states before and after error correction, one expects $F_k\simeq 1$ for $k$ small, while for large $k$, $F_k$ will be small, corresponding to logical errors. The variance of $F_k$ within the small $k$ or large $k$ strata is thus significantly smaller than across all configurations, something that could not be easily recognized or exploited in the earlier unstratified approach of Ref.~\cite{bennink2017} blind to the $k$ value. 

In addition, we mention two further Monte Carlo techniques that gave significant time savings in our examples:
\begin{enumerate}
\item \underline{Self-normalized importance sampling}. For easier numerical implementation, we replace the standard estimator of Eq.~\eqref{eq:FkMk} by the ``self-normalized importance sampling estimator",
\begin{equation}
(\FSN)_{M_k}\equiv \frac{(\widehat F_k)_{M_k}}{(\widehat w_k)_{M_k}},
\end{equation}
where $(\widehat F_k)_{M_k}$ is the estimator of Eq.~\eqref{eq:FkMk} computed from $M_k$ samples, and $\widehat w_k\equiv \frac{1}{M_k}\sum_{m=1}^{M_k}w(\veci_\veca^m)$ is the sample-averaged weight computed from the same $M_k$ samples. This ratio of estimators results in an estimator that depends only on the signs---and not the magnitudes---of the $r_i$s, giving better numerical stability. The self-normalized estimator is biased, but the bias can be estimated from the samples themselves. Regardless, if the bias is observed to be too large, one can always revert back to the unbiased $\widehat F_k$ estimator. In our examples, though the speed of the simulation was already significantly improved by the stratification, the self-normalized estimator did give further variance reduction (this need not hold in other situations). See further details in App.~\ref{app:SamplingSNIS}.

\item \underline{Rejection sampling}. In many settings, we are interested in a range of noise strengths. In this case, rejection sampling can be used, on top of stratified sampling, to reduce the number of stabilizer circuit simulations needed as the noise strength varies. Rejection sampling, like importance sampling, allows us to obtain samples according to a target distribution using samples drawn from a different distribution---a ``reference distribution''. Rather than compensating for the different distribution with weights as in importance sampling, in rejection sampling, we accept or reject reference samples according to the ratio of the target and reference distributions; the accepted samples will then be distributed according to the target. 

As explained in App.~\ref{app:RejectionSampling}, we generate reference samples through stabilizer simulation, for some chosen reference value of $\gamma$. We then use rejection sampling to get samples for other values of $\gamma$. For rejection sampling to outperform direct stabilizer simulation, the acceptance probability has to be large so that we do not need to generate a significantly larger reference pool. This happens when the $\gamma$ values are not too far from the reference. As we see in our examples below, we often have to rely on a few different reference distributions to cover the full $\gamma$ range of interest. Nevertheless, we found rejection sampling to be a big time-saver, compared to full stabilizer simulation for all $\gamma$ values.

\end{enumerate}

\section{Examples}\label{sec:Examples}

The true test of any sampling method is its speed and accuracy in actual problems. We begin with the simple example of the 7-qubit Steane code explored earlier in Ref.~\cite{bennink2017}, as a verification of our stratified sampling approach. Then, we tackle the hardware-relevant problem of assessing the performance of rotated planar surface codes in the presence of general noise. We show the possibility of generating data---with drastically improved runtimes compared with the earlier approach of Ref.~\cite{bennink2017}---for the hardest case of coherent errors for codes as large as distance-15 ($d=15$), involving $2d^2-1=449$ qubits. At the end, to demonstrate applicability of our approach beyond error-correction settings, we describe the simulation of an error-reduction protocol. Further technical details of our sampling approach applicable to all three examples can be found in App.~\ref{app:SimDetails}.

\subsection{7-qubit Steane Code}

We begin with a small-sized example, the 7-qubit Steane code. We follow the protocol described in Ref.~\cite{bennink2017}, for direct comparison with their results on depolarizing and amplitude-damping noise. The example is also small enough for a state-vector exact simulation of the circuit for the case of coherent errors, which provides the ground-truth comparison for our stratified sampling.

The Steane-code protocol in Ref.~\cite{bennink2017} begins with a noiseless encoding (Fig.~5 of \cite{bennink2017}) of seven physical qubits into one of the six single-logical-qubit Pauli eigenstates $\{\ket{0}_L, \ket{1}_L, \ket{\pm}_L, \ket{\pm i}_L\}$. The encoded state then undergoes an idle timestep (i.e., identity operation), during which each physical qubit experiences identical, independent single-qubit noise $\cE$ (Fig.~6 of \cite{bennink2017}). This is followed by three rounds of noisy syndrome extraction, each round comprising the six syndrome measurement circuits specific to the Steane code, with noise $\cE$ added at single-qubit locations throughout the circuits (Figs.~7-12 of \cite{bennink2017}). One additional round of noiseless syndrome extraction and recovery is performed at the end to remove remaining correctable errors on the output, to distinguish between a genuine logical error on the output versus remnant correctable errors that can be removed by further error correction rounds. The fidelity between the output logical state and the input state is computed (the $O$ in $F$ is thus the input logical state). The average infidelity (deviation of fidelity from unity) over the six input logical states is computed and plotted against the physical channel infidelity $\epsilon$, calculated as the worst-case (maximum) infidelity over all input states for $\cE$.

\begin{figure*}
\includegraphics[trim=16mm 115mm 16mm 31mm, clip, width=\textwidth]{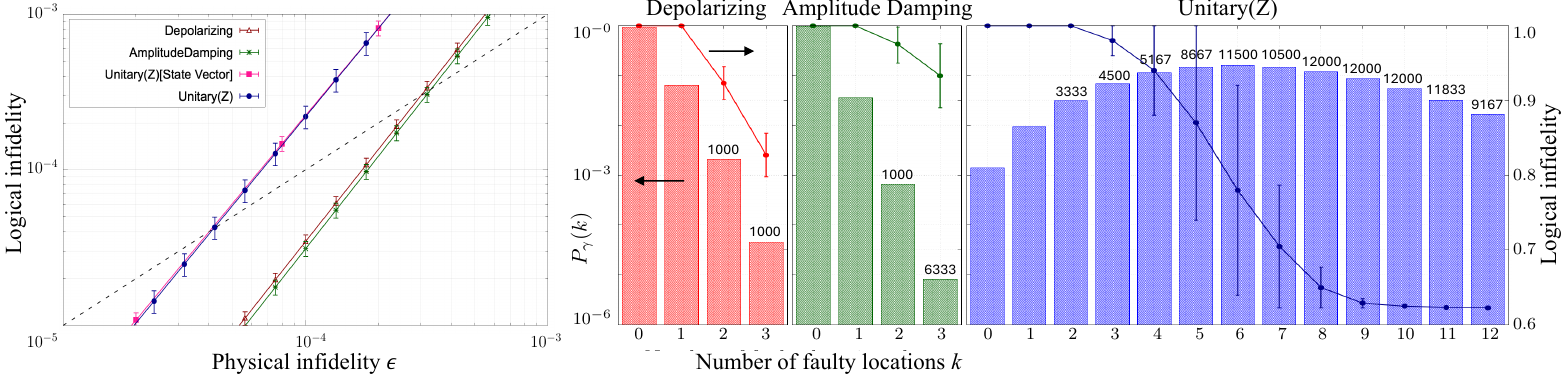}
 \caption{\label{fig:SteaneCode} The 7-qubit Steane code under depolarizing, amplitude-damping, and $Z$-rotation coherent [Unitary(Z)] noise. Leftmost plot: Logical infidelity computed using our stratified sampling approach, together with an exact state-vector simulation for the $Z$-rotation noise. Right histograms and plots: The histograms give $P_\gamma(k)$ for the $k$ values with significant probability of occurrence for $\epsilon_\textrm{ref}$; the lines give the logical fidelity $F_k$ value for given $k$, with the error bars and number of samples (numerical values at the top of each histogram bar) to achieve a 10\% confidence interval around the estimated $F$ value with a 99\% confidence level.}
\end{figure*}

Ref.~\cite{bennink2017} presented results for $\cE$ as the single-qubit depolarizing (D) and amplitude-damping (A) channels,
\begin{align}\label{eq:EDEA}
\cE_\textrm{D}(\,\cdot\,)&\equiv(1-p_\textrm{D})(\,\cdot\,)\\
&\qquad+\tfrac{1}{3}p_\textrm{D}{\left[X(\,\cdot\,)X+Y(\,\cdot\,)Y+Z(\,\cdot\,)Z\right]},\nonumber\\
\cE_\textrm{A}(\,\cdot\,)&\equiv E_0(\,\cdot\,)E_0^\dagger +E_1(\,\cdot\,)E_1^\dagger,\nonumber\\
\textrm{with }\quad E_0&\equiv \ketbra{0}{0}+\sqrt{1-p_\textrm{A}}\ketbra{1}{1}, E_1\equiv \sqrt{p_\textrm{A}}\ketbra{0}{1},\nonumber
\end{align}
with $p_D,p_A\in[0,1]$ (usually $\ll 1$). $\cE_\textrm{D}$ is already in the form of a stabilizer-channel decomposition of Eq.~\eqref{eq:WeakNoise}, with $\gamma\equiv p_D$ while $\cF\equiv \frac{1}{3}(\cX+\cY+\cZ)$ composed of the three Pauli channels, so $\nu=0$ (negativity of $\cF$). $\cE_\textrm{A}$ can be written in a similar form [Eq.~(6) of \cite{bennink2017}], with $\gamma\equiv \frac{1}{2}{\left[(1+p_\textrm{A})-\sqrt{1-p_\textrm{A}}\right]}=\frac{3}{4}p_\textrm{A}+O(p_\textrm{A}^2)$, and $\cF\equiv \frac{1}{2\gamma}[(1-p_\textrm{A})-\sqrt{1-p_\textrm{A}}]\cZ+\frac{1}{\gamma}p_\textrm{A}\cR_Z=-\frac{1}{3}\cZ+\frac{4}{3}\cR_Z+O(p_\textrm{A})$, so $\nu=\frac{1}{3}$. Here, $\cR_Z(\,\cdot\,$ is the channel corresponding to a reset in the $\ket 0$ state (i.e., measure in the $Z$ basis, and do a bit-flip if the $-1$-eigenstate outcome is obtained). The physical channel infidelity $\epsilon$ depends in a different way on $\gamma$ for each channel.

Using our stratified sampling, we simulated the Steane-code protocol for both channels. Further simulation details are given in App.~\ref{app:SimDetails} (the same protocol is used in all our examples); the results are shown in Fig.~\ref{fig:SteaneCode} (left plot). We see very good agreement between the results from our simulation and those found in Ref.~\cite{bennink2017} (see Figs.~3 and 4 in that reference). For further verification of our sampling accuracy, we also simulated a unitary $\cE$, specifically, the single-qubit $Z$ rotation, $\cE(\,\cdot\,)=Z_\theta(\,\cdot\,)Z_\theta^\dagger$, where $Z_\theta=\ketbra{0}{0}+\upe^{\upi\theta}\ketbra{1}{1}$ is unitary. The choice of a unitary noise example here is two-fold: The unitary nature entails that the quantum state remains a pure state even in the presence of noise; this allows the use of state-vector quantum simulation for exact calculation of the infidelity values. In addition, this unitary case was difficult for the standard importance sampling method of Ref.~\cite{bennink2017}, but our stratified sampling gives easy access. Figure \ref{fig:SteaneCode} shows excellent agreement between the exact simulation results and the data from stratified sampling. 

It is useful to compare the relative costs of simulating the different noise channels for this small-sized Steane-code example. For each channel here, a single reference distribution---at reference physical infidelity value $\epsilon_\textrm{ref}=2\times 10^{-4}$---sufficed to cover the full range of $\epsilon$ between $10^{-5}$ and $10^{-3}$; samples for other $\epsilon$ values were generated from the reference pool with rejection sampling (acceptance probability $>95\%$), which took negligible time compared with that for the full circuit simulation needed to generate the reference samples. We thus only discuss the cost of generating those reference samples. 

The histograms (with overlaid plots) in Fig.~\ref{fig:SteaneCode} show the sampling data at $\epsilon=\epsilon_\textrm{ref}$, with the respective $\gamma$ values for each channel: The histograms give the relevant $k$ values (i.e., $\cK$), chosen according to $P_\gamma(k)$; the overlaid plots give the fidelity estimates for each $k$, with the number of samples for each $k$ needed to achieve the target precision given at the top of each histogram bar; see App.~\ref{app:SampleAllocation} for further details. Three observations are in order: First, the relevant $k$ values are those with large enough \emph{product} of $P_\gamma(k)$ and $F_k$ to contribute to the overall $F$ value; ones with too-small $P_\gamma(k)$ and/or too-small $F_k$ simply do not matter. Second, the $F_k$ error bars only need to be small enough for those errors to not contribute to the error in the overall $F$; in practice, we increase the sample sizes across $k$ as we gather more data, until we find that the $F$ value stabilizes. Third, the $Z$-rotation noise involves a much larger set of relevant $k$ values compared with depolarizing or amplitude-damping noise, consistent with our expectation that it is harder to simulate, and reminiscent of the earlier difficulties in using randomized benchmarking for coherent errors \cite{wallman2014}.

The per-sample full-circuit simulation time is about the same across the three channels considered here and across strata (determined largely by the size of the Steane-code circuits), so the simulation cost is proportional to the total sample size needed for the desired estimate precision. The depolarizing channel needed 2000 samples in total across all relevant strata; the amplitude-damping channel required 7333 samples; the unitary-$Z$ channel needed just under $101,000$ samples. The nonunitary amplitude-damping channel required less than 4$\times$ the samples for the depolarizing channel, a behavior we will observe again in our surface-code examples. The unitary channel needed significantly more samples, about 50$\times$ that for the depolarizing channel. This may appear large, but one should remember that this is a drastic improvement over the earlier method of Ref.~\cite{bennink2017} which never converged to any reasonable precision with even larger sample sizes. Furthermore, we see from the fidelity plots that one need not simulate the unitary situation for all relevant $k$ values---the smooth curve allows for easy interpolation using only a fraction of the relevant $k$ values. For this small-sized example, we did not do such an interpolation; this, however, becomes an important time-saver in our surface-code example below. 

To give a sense of the actual time taken, one sample with full-circuit simulation (i.e., for the reference) for the Steane-code protocol took about 15$\mu$s on a standard desktop, so the depolarizing, amplitude-damping, and unitary-$Z$ simulations for a single $\gamma$ value took total time $\sim$0.03s, 0.1s, and 1.5s, respectively.

\subsection{Rotated planar surface codes}

\begin{figure}
\includegraphics[trim=8mm 165mm 7mm 50mm, clip, width=\columnwidth]{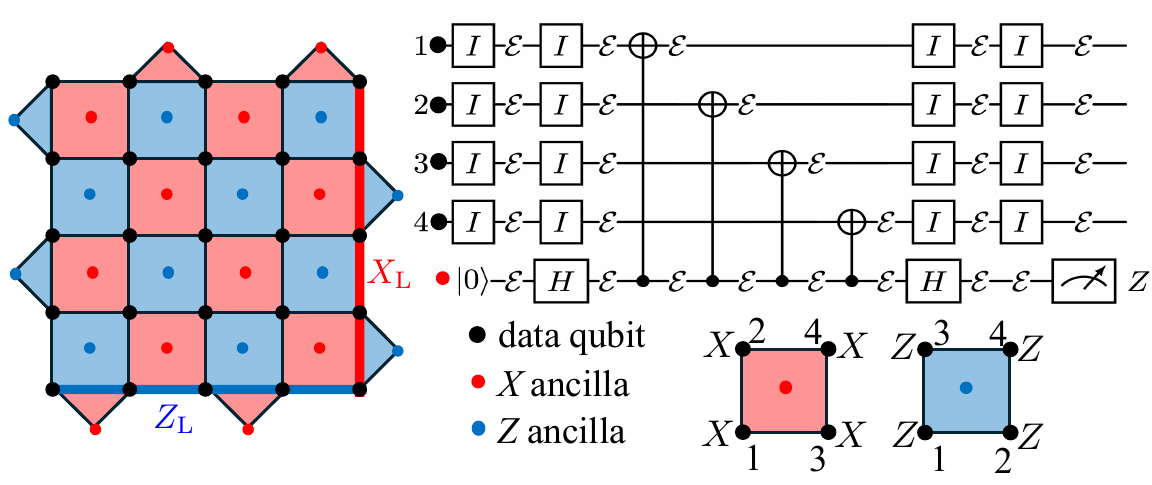}
\caption{\label{fig:SurfCodePatch} Rotated planar surface code. The checkerboard is a surface-code patch for $d=5$, encoding one logical qubit. Red(Blue) squares (or triangles) are for the $X(Z)$ stabilizers. The logical $X(Z)$ operator is the $X(Z)$ operator on every physical qubit along a single column(row), as indicated by the red(blue) vertical(horizontal) line. The circuit shown measures the $XXXX$ stabilizer for detecting $Z$ errors, with the qubit labels 1--4 referring to the four data qubits around the ancillary qubit. Noise $\cE$ is inserted at the indicated locations. The corresponding circuit measuring the $ZZZZ$ stabilizer for detecting $X$ errors, done synchronously, is identical, except that the target and control qubits are swapped for CNOTs and the two Hadamard $H$ gates on the ancillary qubit are replaced by identity gates; the qubit labels 1--4 also refer to the data qubits around the $Z$-ancillary qubit in a different order. Data qubits in the bulk participate in CNOT gates in one of these circuits in every CNOT time step; those on the boundaries have additional identity gates---with insertions of $\cE$---if they participate in no CNOT gate in that time step.}
\end{figure}

Next, we turn to our main example, tackling the popular rotated planar surface code in the presence of general noise. The rotated planar surface code---or surface code for short---comprises a checkerboard arrangement of local $X$ and $Z$ stabilizer generators, with data qubits located at the vertices of the lattice; see Fig.~\ref{fig:SurfCodePatch}. Syndrome measurement ancillary qubits are at the centre of each square, and at specific locations along the boundary. Each $d^2$ ($d$ an odd integer) lattice of data qubits encode one logical qubit, and permits correction of arbitrary errors on up to $t\equiv\frac{1}{2}(d-1)$ data qubits, assuming noise-free syndrome measurements. Together with $d^2-1$ ancillary qubits for the syndrome measurements, we have $2d^2-1$ physical qubits in all. Each ancillary qubit measures a single stabilizer generator, carried out via four CNOTs that connect the surrounding four data qubits to the ancillary qubit, followed by a measurement of the ancillary qubit. These syndrome-measurement operations, in reality, can also have errors. By repeating the syndrome measurement $d$ times and analyzing the $d$-cycle syndromes together, however, we retain the ability to correct errors on $t$ or fewer qubits, including the ancillary qubits. 

We consider a standard memory experiment: The surface-code qubits undergo $d$ cycles of syndrome measurements across the entire lattice in the presence of circuit-level noise. The syndromes are decoded after the $d$ cycles (using PyMatching \cite{higgott2022} with the Sparse-Blossom algorithm \cite{higgott2025}), and any requisite virtual correction is applied. Figure \ref{fig:SurfCodePatch} (top right) shows the precise circuits employed. Single-qubit noise $\cE$ is inserted after every state preparation and gate (including identity gate, i.e., waiting time); $\cE$ is inserted right before the final $Z$ measurement. Our circuit-level noise assumes identical and independent single-qubit noise $\cE$ everywhere, including at the two-qubit CNOT locations. In reality, a two-qubit gate can cause genuine two-qubit correlated noise, and the noise can vary across circuit locations, both of which can be accommodated by our stratified sampling approach. Nevertheless, we restrict here to single-qubit homogenous noise so that we have a single noise $\cE$ everywhere.

The above describes the noisy circuit $\widetilde\cir$. In addition, we need the observable $O$ and the input state $\rho$ to form the quantity of interest $F=\tr\{O\widetilde\cir(\rho)\}$. The relevant figure of merit here is the worst-case---over all input code states---fidelity between the output and input states of the memory experiment. This worst-case fidelity can be computed, not by running the experiment for all input logical qubit states and then taking the minimum, but by estimating $F$s using stratified sampling for a discrete (small) number of $\rho$s and $O$s, and then doing the minimization in post-processing. For that, observe that we can compute, via linearity, $F(\rho,O;\widetilde\cir)$ for any \emph{logical} $\rho_\mathrm{L}$ and $O_\mathrm{L}$ from the values $F(i,j)\equiv F((\varrho_\mathrm{L})_i,(\varrho_\mathrm{L})_j;\widetilde\cir)$ for all $(\varrho_\mathrm{L})_i$, $(\varrho_\mathrm{L})_j$ logical stabilizer states (i.e., eigenstates of the logical $X$, $Y$, and $Z$ operators) which form a basis for the logical qubit space. For fidelity between input and output code states, $\rho=O=\rho_\mathrm{L}$. Thus, we compute $F$ for any $\rho_\mathrm{L}$ using $F(i,j)$s estimated using our sampling protocol, and then minimize over $\rho_\mathrm{L}$ for worst-case fidelity. 

\begin{figure*}
\includegraphics[width=0.85\textwidth]{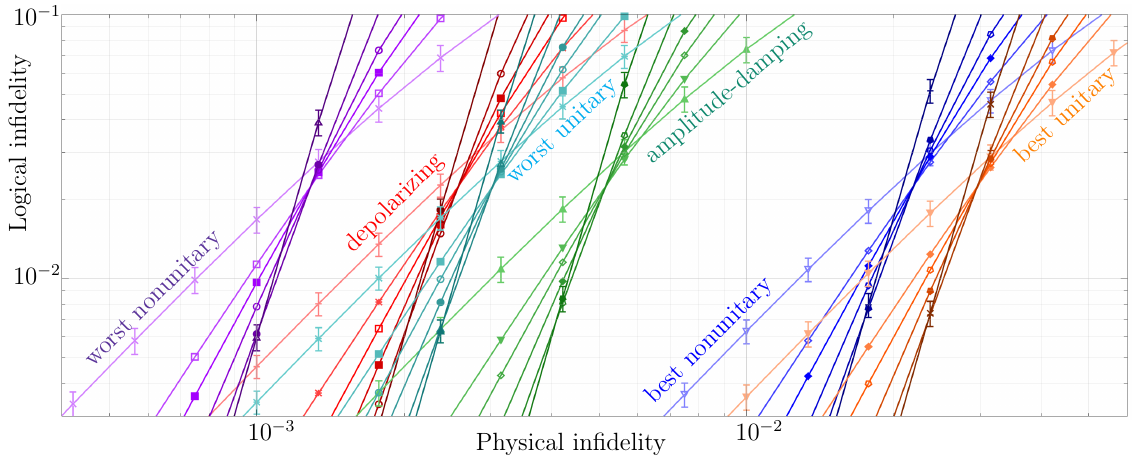}
 \caption{\label{fig:SurfCodeAll} Logical infidelity versus physical infidelity for different noise channels, for the rotated surface code of code distances $d=3,5,7,9,11,15$, involving, respectively, $2d^2-1=17,49, 97,161, 241,449$ qubits. The lines with the smallest gradients are for $d=3$; the line gradients increase monotonically with $d$. The different colored groups of lines correspond to different noise channels: depolarizing noise (red), amplitude-damping noise(green), worst/best (i.e., largest/smallest logical infidelity for given physical infidelity) random non-unitary noise (purple/blue), and worst/best random unitary noise (cyan/orange). The error bars, corresponding to 10\% confidence interval with a 99\% confidence level, are shown only for the smallest and largest $d$ values; intermediate $d$ values have similar error-bar sizes.}
\end{figure*}

We simulate a variety of single-qubit noise channels. To make contact with past surface-code simulations, we include the depolarizing channel [Eq.~\eqref{eq:EDEA}]. We also examine amplitude-damping noise. For general noise, we randomly choose unitary and non-unitary CPTP $\cE$. For nonunitary channels, for $J=2,3,\ldots,5$, we randomly generate $(J-1)$ $2\times 2$ matrices (in the computational basis), $\widetilde{E}_1, \ldots, \widetilde{E}_{J-1}$. Each $\widetilde{E}_j$ has complex entries, with real and imaginary parts independently sampled from a standard normal distribution. Then, the random non-unitary noise $\cE$ is constructed as $\cE(\,\cdot\,)\equiv \sum_{j=1}^J E_j(\,\cdot\,)E_j^\dagger$, with $E_j\equiv \sqrt\omega \widetilde E_j$ for $j=1,\ldots, J-1$, and $E_J\equiv \sqrt{\id-\sum_{j=1}^{J-1}E_j^\dagger E_j}$. Here, $\omega\ll1$ is a positive constant used to control the noise strength, while ensuring $\id-\sum_{j=1}^{J-1}E_j^\dagger E_j\geq 0$.
For unitary (or coherent) noise, we write $\cE_U(\,\cdot\,)=U(\,\cdot\,)U^\dagger$, with $U\equiv \upe^{\upi\delta\widehat{\boldsymbol{n}}\cdot\boldsymbol{\sigma}}$, for $0\leq \delta\ll 1$ the noise strength, $\boldsymbol{\sigma}\equiv (X,Y,Z)$ the vector of Pauli operators, and $\widehat{\boldsymbol{n}}\equiv (\sin\theta\cos\phi,\sin\theta\sin\phi,\cos\theta)$ a unit vector with $\theta$ and $\phi$ uniformly randomly chosen from $[0,\pi)$ and $[0,2\pi)$, respectively. 

Figure \ref{fig:SurfCodeAll} shows the simulation results. For each chosen $\cE$, the noise strength $\gamma$ ($=\gamma_\mathrm{dp}$, $\gamma_\mathrm{ad}$, $\omega$, or $\delta$) is varied over a range. The stabilizer decomposition is done for each noise-strength value, choosing the one with the smallest negativity. The worst-case logical infidelity is computed for each noise-strength value. This is plotted on the vertical scale (``Logical infidelity") against the physical channel infidelity $\epsilon$. We only show the lines for the best and worst logical infidelities for the $d=3$ case across the non-unitary random channels, and separately across the unitary ones; based on the $d=3$ performance, we simulate higher-distance codes with those ``best" and ``worst" random channels. The lines on the plot show the estimated logical infidelity, with the error bars indicated for each data point. Stratified sampling is run for enough samples until the error bars are similar in size across the range of physical infidelity values and noise channels. 

Figure \ref{fig:SurfCodeAll} highlights the strong dependence of the noise threshold---the value of physical infidelity below which increasing the code distance decreases the logical infidelity---on the specific noise channel, with threshold numbers that vary over more than an order of magnitude as we vary the noise type. The depolarizing noise (red lines in the figure) result, which reproduces the often-quoted $10^{-3}$ noise threshold value from standard stabilizer simulations, tells only a small part of the actual story. Interestingly though, it gives a threshold stringent enough to cover almost all noise---the true threshold, over all noise, is the one given by the worst-non-unitary channel we found, and that is only about half the threshold value predicted by just looking at depolarizing noise alone. Unitary noise seems to not be especially adversarial for surface codes, with the worst unitary noise we found (cyan lines) having performance quite similar to that from depolarizing noise. This extends and confirms the conclusions of the earlier work of Ref.~\cite{bravyi2018} with similar findings for the restricted case of $Z$-rotation errors in the code-capacity (i.e., input only, not circuit-level, noise) setting. The worst-performing channel was found to be a nonunitary channel; the channel is given in App.~\ref{app:WorstE}. It would be interesting to understand why this gave the worst performance. A notable feature is that it has a larger proportion of Pauli-reset channels of different types, compared to most other channels tested.

\begin{figure}
\includegraphics[width=\columnwidth]{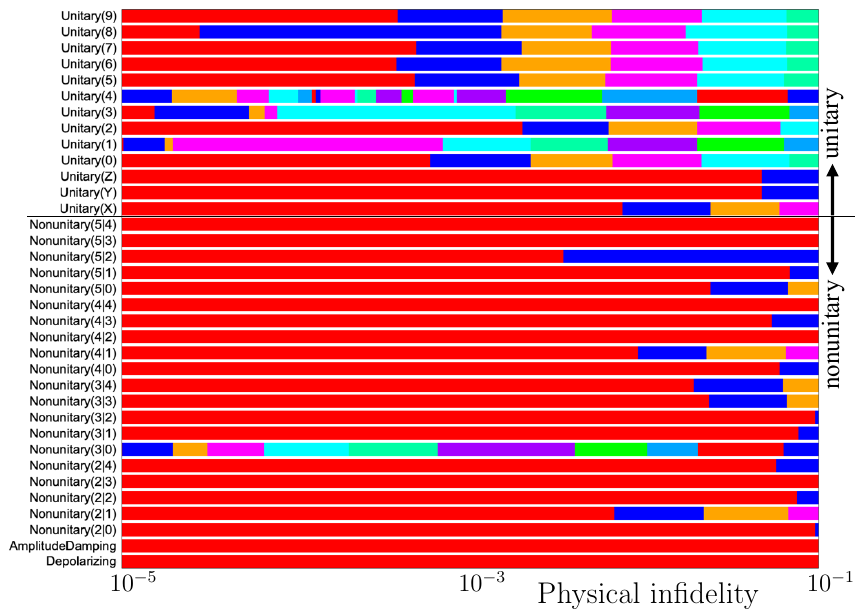}
 \caption{\label{fig:RefRanges} Reference ranges for rejection sampling, for $>95\%$ acceptance probability, across the physical channel infidelity range of interest. Each horizontal bar corresponds to a single noise channel: unitary channels in the top half (with unitary-$Z$, $Y$, and $X$ channels as the bottom three); nonunitary channels in the bottom half (with amplitude-damping and depolarizing channels as the bottom two). Each colored segment uses a single reference distribution for rejection sampling. The channels marked Unitary($\ell$), $\ell=0,1,\ldots 9$ are randomly generated, as are the nonunitary channels marked Nonunitary$(J|\ell)$, with $J$ Kraus operators; see main text for the channel generation procedure.}
\end{figure}

Now, let us examine the simulation cost of this surface-code example. As in the Steane-code example, the per-sample time for full-circuit simulation is similar across different noise channels, though it does increase with code distance: Typical times are (in ms) 0.03, 0.08, 0.3, 1.2, 3.5, and 20, for $d=$3, 5, 7, 9, 11, and 15, respectively. Again, rejection sampling times are negligible compared to the circuit simulation time for generating reference samples, so we only need to compare the reference sample sizes deduce the relative costs of simulating the different noise channels.

For that, a major contributing factor is the number of reference distributions needed. As the physical infidelity $\epsilon$ is varied, for the same noise channel, one may need multiple reference distributions, to cover the full range with sufficiently high acceptance probabilities. Fig.~\ref{fig:RefRanges} shows the number of reference distributions used for a variety of noise channels. For our $>95\%$ acceptance probability criterion for rejection sampling, many channels require more than one reference distribution. Unitary channels have a higher occurrence of high-number of reference distributions, with most needing 6--9 of them; the extreme case required 20 different ones. Most nonunitary channels need 1--4 reference distributions, though there is a single example with 11 distributions (different from the channel that gave the worst logical infidelity). One can use fewer reference distributions to cover the full $\epsilon$ range with a less stringent acceptance probability, at the cost of having to generate larger sample pools for each reference---this is an optimization that can be done for each noise channel. One can also optimize reference ranges based on the number of faults $k$. Here, for better comparison across all the noise channels, we made an \emph{a priori} conservative choice of $>95\%$ acceptance probability, without any per-channel or per-$k$ optimization.

Let us examine then the cost of generating the samples for each reference distribution. As in the Steane-code example, for each channel, each code distance $d$, as well as each reference infidelity value $\epsilon_\textrm{ref}$, different $k$ values are relevant. In practice though, we observe the relevant $k$ range to be quite independent of the $\epsilon_\textrm{ref}$ values. We also mentioned earlier that the fidelity varies smoothly over $k$, so we need not sample all relevant $k$s but only critical ones that allow for a good interpolation for the fidelity. For that, we make assumptions on the fidelity values for small and large $k$: For small $k$ below the code correction capability (i.e., $k\leq t\equiv (d-1)/2$), we assume the decoder correctly removes the error so the logical fidelity is 1; for large $k$ far beyond the correction capability of the code, the logical state is assumed to be completely random (i.e., the completely mixed qubit state), so that the logical fidelity is $0.5$. In between those two extremes, we need careful sampling especially in the critical zone where the fidelity starts to fall from $1$. By starting with small and large $k$ values, and then gradually homing in on those in-between critical $k$ values, we do not actually need many samples to get a sufficiently precise fidelity estimate. Appendix \ref{app:Interpolate} gives further details on this interpolation.

As an example, the sampled $k$ values and the corresponding sample sizes $M_k$s for the $d=7$ case are given in Tab.~\ref{tab:SurfCodeData} in the appendix, with the accompanying Fig.~\ref{app:SurfCodeData} showing the smoothly varying fidelity curves. Again, we see that the number of relevant $k$ values is generally larger for the unitary channels than for other channels, and the nonunitary channels can have similar number of relevant $k$s as the depolarizing channel. However, because the critical zone of the fidelity is observed to be quite similar in extent across the different channels, the actual number of $k$ values we sample for good interpolation is not very different (a factor of 2 or 3) between unitary and nonunitary channels.

\begin{table}
\begin{tabular}{c|cccc}
&$M$&No.~of $\epsilon_\textrm{ref}$s&Time per $\epsilon_\textrm{ref}$ (s)&Total time (s)\\
\hline\hline
Depol&7,000&1&2.2&2.2\\
\hline
AD&10,000&1&3.1&3.1\\
\hline
$NU_\textrm{best}$&14,000&1&4.3&4.3\\
\hline
$NU_\textrm{worst}$&8,500&2&2.6&5.2\\
\hline
$U_\textrm{best}$&44,000&6&13.6&81.6\\
\hline
$U_\textrm{worst}$&24,500&6&7.6&45.6
\end{tabular}
\caption{\label{tab:d7Timing} Resource costs for the $d=7$ surface code, to generate samples for the reference values $\epsilon_\textrm{ref}$s for the six channels in Fig.~\ref{fig:SurfCodeAll}: depolarizing (Depol), amplitude-damping (AD), best nonunitary ($NU_\textrm{best}$), worst nonunitary ($NU_\textrm{worst}$), best unitary ($U_\textrm{best}$), and worst unitary ($U_\textrm{worst}$) channel). $M$ is the total reference samples needed per $\epsilon_\textrm{ref}$ for the target precision, and the number of $\epsilon_\textrm{ref}$s  refers to the number of reference infidelity values needed to cover the range $\epsilon\in[10^{-5},10^{-1}]$ in our example. The total time (last column) is the product of the number of $\epsilon_\textrm{ref}$s and the time per $\epsilon_\textrm{ref}$, giving the upper bound on the total simulation time; in practice, one often does not need all $\epsilon_\textrm{ref}$ values.}
\end{table}

To give a sense of the actual time costs for the simulation, Tab.~\ref{tab:d7Timing} lists the total sample sizes and time taken for generating the reference pools for the $d=7$ case; Table~\ref{tab:Mvalues} in the appendix gives the data for other $d$ values, which show very similar behavior in relative costs between channels. From Tab.~\ref{tab:d7Timing}, we see that, to generate samples for a single reference value, the slowest case (one of the unitary channels) takes but about six times that for the depolarizing channel; the nonunitary ones have very comparable times as for the depolarizing channel. Now, the total time taken to cover the whole infidelity range depends also on the number of reference values needed; that total time is given in the rightmost column of Tab.~\ref{tab:d7Timing}. We then see the unitary channels costing significantly more time ($\lesssim$ a factor of 40 compared with the depolarizing channel). Nevertheless, to produce lines like those in Fig.~\ref{fig:SurfCodeAll}, one does not need to go over the full range of $\epsilon$ values---having a few data points often suffice. We can then strategically pick those to be the $\epsilon$ values covered by a subset of the $\epsilon_\textrm{ref}$s values, thereby reducing the actual cost of simulation, so that the unitary channels become nearly as cheap as the nonunitary ones.


\subsection{Clifford circuit noise reduction}

As a final example to illustrate the applicability of our stratified sampling procedure beyond the error-correction setting, we examine a recent noise-reduction protocol for Clifford circuits described in Ref.~\cite{delfosse2025}. The Clifford noise reduction---or ``CliNR''---protocol takes one step beyond error mitigation towards error-corrected operation by adopting a teleportation approach---standard in fault-tolerant quantum computing schemes---to implement Clifford gates. This allows offline fault detection in the ancillary states, to assure low error rates, before connecting those states to the computer. We refer the reader to the original article for a full discussion of the CliNR protocol. Here, we provide only the details necessary to understand our simulation results.

A given $n$-qubit Clifford circuit $\cir$ with $s$ gates is partitioned into $t$ subcircuits, with $t$ chosen to satisfy a gate-overhead constraint, $\omega_G \leq \omega_G^{\max}$ (e.g., $\omega_G^{\max}=2$ or $4$). Each subcircuit is applied to one half of $n$ Bell-state pairs of ancillary qubits, and then $r$ random stabilizer measurements are performed on the ancillary qubits to check for errors, with $r=\left\lfloor \log\left(\frac{s}{n}\right) \right\rfloor$. If any measurement yields a nontrivial outcome, the protocol is restarted. Only when all measurement results are trivial do we perform the teleportation by connecting the other half of the Bell-state pairs to the computational qubits, thereby applying $\cir$ on the computational qubits. 

The noise model used in the original work appends noise after every physical operation: Single-qubit gates and idle steps are followed by a depolarizing channel with error rate $p_1$, while two-qubit gates are followed by a two-qubit depolarizing channel with error rate $p_2$. To check our implementation of the CliNR protocol, we first used this noise model in our simulations to reproduce the reported graphs in Ref.~\cite{delfosse2025}.

\begin{figure}
\includegraphics[width=\columnwidth]{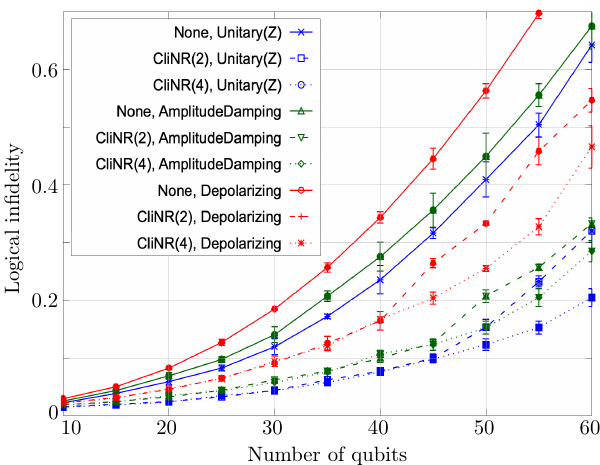}
\caption{\label{fig:CliNR} The CliNR protocol of Ref.~\cite{delfosse2025} under depolarizing, amplitude-damping, and unitary-$Z$ rotation noise, simulated using stratified sampling. For each noise, we plot the logical infidelity for the target circuit $\cir$ against the number of qubits $n$ involved in $\cir$, for three scenarios: no fault-detection protocol (`None'), CliNR with $w_G\leq 2$ [`CliNR(2)'], and CliNR with $w_G\leq 4$ [`CliNR(4)']. For all noise channels, the physical channel (worst-case) infidelity is $10^{-4}$.}
\end{figure}

Next, we use stratified sampling to explore the performance of CliNR beyond the depolarizing noise discussed in the original work. As in earlier examples, we measure noise by---logical or physical---channel infidelity, and we apply only independent single-qubit noise, even for two-qubit gates. Since, for the depolarizing channel, the channel infidelity is proportional to $p$, we replace all $p$s in the expressions in Ref.~\cite{delfosse2025} for the protocol parameters by the physical channel infidelity. 
Our stratified sampling simulation results are presented in Fig.~\ref{fig:CliNR}, for depolarizing, amplitude-damping, and unitary-$Z$ noise. Again, we observe here that the unitary noise is not particularly adversarial for the CliNR protocol, with depolarizing noise giving the worst logical infidelity among the three noise types. Note that the kinks in the plot lines are due to discrete switching of the protocol parameters.

\section{Discussion and Conclusions}\label{sec:Conc}

We have described how stratified importance sampling, in conjunction with rejection sampling, can give access to general noise in stabilizer simulations of Clifford circuits. The stratification according to number $k$ of faulty circuit locations is natural in the weak-noise setting relevant for quantum computing contexts. It allows us to zoom in on the relevant $k$ values, and do a fine-grained sampling that allocates sample sizes according to the variability of the quantity of interest for given $k$. This resolved the too-large variance issues that led to the failure of previous methods like that in Ref.~\cite{bennink2017}. The stratification in $k$ further allowed interpolation shortcuts for $F$s that vary smoothly with $k$, giving significant savings in runtime. We were thus able to simulate general noise, including the difficult case of unitary or coherent errors, in our examples in reasonable time. In particular, we observe that nonunitary noise comes at simulation costs comparable with that for Pauli noise; unitary noise requires more samples, but the estimates still converge in reasonable time even for the large $\sim$450-qubit surface-code example.

General noise can incur greater sampling costs, compared with Pauli noise, in two ways, by having a larger set of relevant $k$ values, and in requiring more reference distributions to cover the full physical infidelity range of interest. We note, however, that there is a situation for which a single reference distribution always suffices: when the relative weights between different possible errors at each location is noise-strength independent, i.e., $r_i^{(a)}$s of Eq.~\eqref{eq:WeakNoise} are independent of $\gamma^{(a)}$. In this case, the only $\gamma$ dependence is in $P_\gamma(k)$; within each $k$ stratum, there is no $\gamma$ dependence, and $F_k$s are hence $\gamma$-independent. We note that standard Pauli noise, including asymmetric (i.e., not just depolarizing) Pauli channel, falls into this category. In such cases, with the samples from a single reference value of $\gamma$, we can access all other values of $\gamma$ simply by weighting each $F_k$ value by $P_\gamma(k)$ for the appropriate $\gamma$ value. Now, of course, to get estimates within a certain fractional precision over all $\gamma$ values, we would need to have sufficiently many reference samples, but since the same set of samples is useful for all $\gamma$ values, we can invest all available computational resources to building up that single reference pool. 

We considered only single-qubit noise channels in our examples. In real hardware, two-qubit noise---or even multi-qubit noise---is often important, arising from imperfections in the gate operations. The stabilizer decomposition [Eq.~\eqref{eq:StabDecomp}] in fact works for general $n$-qubit CPTP channels, but of course, with increasing $n$, the potential number of stabilizer channels in the decomposition rapidly increases (e.g., there are more than 11 thousand two-qubit Clifford gates, and 92 million three-qubit Clifford gates). The hope is that relevant multi-qubit noise are well-approximated by channels with short stabilizer description, so that we do not need to access the whole variety of Clifford errors. It will be useful to investigate this for experimentally relevant two-qubit noise.

In our examples, we explored infidelity values no smaller than $10^{-5}$. For surface codes, this is already well below the fault-tolerance threshold value, and the logical infidelities for smaller physical infidelities are expected follow well a straight line that can be extrapolated from larger physical infidelity values. Nevertheless, it might be useful to directly simulate for small infidelity values, e.g., for the latest studies on good q-LDPC code families with much higher minimal code distances and hence much smaller logical infidelity values. In this small infidelity limit, with correspondingly stringent fractional precision requirements, stratified sampling may not work sufficiently well to handle these ``rare-event" scenarios. This was pointed out in a very recent paper \cite{mayer2025}, which extended an earlier ``splitting method" by Bravyi and Vargo \cite{bravyi2013} for rare-event simulation in error-correction settings with Pauli noise; see also the even newer article \cite{beverland2025} addressing the same problem. It would be interesting to see if one could generalize these rare-event methods also to beyond-Pauli noise, perhaps employing again the stabilizer decomposition of general channels.

The efficacy of variance-reduction methods for Monte Carlo sampling generally depends on the problem details. Here, we have shown the usefulness of stratified sampling for three specific scenarios---Steane-code protocol, surface codes, as well as the CliNR protocol, with $F$ taken to be a fidelity metric. We expect it to work well in related contexts, whenever the per-stratum $F_k$ value varies significantly over $k$. We invite the reader to apply our methods, and discover new contexts where such stratification helps.

\section*{Acknowledgments}
We are grateful to Erio Trong Duong and Aleksandr Rodin for helpful discussions. M Myers II is supported by a Centre for Quantum Technologies (CQT) PhD scholarship. This project is supported by the National Research Foundation, Singapore through the National Quantum Office, hosted in A*STAR, under its Centre for Quantum Technologies Funding Initiative (S24Q2d0009). The computational work for this article was partially performed on resources of the National Supercomputing Centre, Singapore (https://www.nscc.sg). The title of our work was inspired by the wonderfully named paper ``Magic state cultivation: growing T states as cheap as CNOT gates" by C Gidney, N Shutty and C Jones [arXiv:2409.17595]. Data and code are available upon request from the authors.

\bibliography{SimGenNoise.bib}

\appendix


\section{Technical details of the stratified importance sampling protocol}\label{app:Sampling}
Here, we provide further details of the stratified importance sampling protocol omitted in the main text.

\subsection{Standard importance sampling estimator}\label{app:SamplingIS}
In the main text, we defined the standard importance sampling estimator
\begin{equation}
\widehat{F}_M = \frac{1}{M}\sum_{m=1}^M w({\vecmu^m})f({\vecmu^m}),
\end{equation}
with $w({\vecmu^m})\equiv \frac{q(\vecmu^m)}{p(\vecmu^m)}$.
Observe that, for any location $a$, the expectation value of $w_\mu^{(a)}$ is unity:
$\Exp[w^{(a)}]=\sum_{\mu_a=0}^{I^{(a)}} p^{(a)}_{\mu_a} w^{(a)}_{\mu_a}
=\sum_{{\mu_a}=0}^{I^{(a)}}q^{(a)}_{\mu_a}=1$.
Since different locations are sampled independently, it follows that $\Exp[w(\vecmu)]=1$.

The estimator $\widehat F_M$ has variance $\Var[\widehat F_M]=\frac{1}{M}\Var[\widehat F_1]=\frac{1}{M}{\left(\Exp[w(\vecmu)^2|f(\vecmu)|^2]-|F|^2\right)}$.
Now, $|F|\geq 0$, and observe that $f$ is a fidelity between two states, the pure state $\varrho_{\mu_{A+1}}$ and $\cG^{(A)}\cS_{\mu_{A}}\ldots \cG^{(1)}\cS_{\mu_1}(\varrho_{\mu_0})$, so $f\in[0,1]$. We assume no knowledge of how $f$ changes with $\vecmu$; otherwise, we can use that knowledge to choose $p$s to minimize the variance. Then,
\begin{equation}\label{eq:VFM}
\Var[\widehat F_M]=\frac{1}{M}{\left(\Exp[w^2|f|^2]-|F|^2\right)}\leq \frac{1}{M}~\Exp[w(\vecmu)^2],
\end{equation}
as stated in the main text.

\subsection{Stratified importance sampling estimator}\label{app:SamplingSIS}
We argued in the main text that the variance of the stratified estimator has significantly smaller variance than the standard importance sampling estimator whenever $F_k$ depends strongly on $k$. We can see this clearly by examining the variance of the stratified estimator: $\Var[(\widehat F_\textrm{str})_M]=\sum_{k\in\cK}P_\gamma(k)^2\frac{1}{M_k}\Var_k[(\widehat F_k)_1]$, with the subscript $k$ on $\Var_k$ indicating that the variance is for the conditional probability distribution $\{p(\veci_\veca|k), \veci_\veca\in\cC_k\}$. Already, the variance is reduced compared to un-stratified sampling by excluding variability from $k\notin\cK$ through the use of prior knowledge of $F_{\notin\cK}$. More importantly, we note that $\Var[(\widehat F_\textrm{str})_M]=\sum_{k\in\cK}\frac{P_\gamma(k)^2}{M_k}\Var_k[(\widehat F_k)_1]\leq \frac{P_\cK}{M}\!\sum_{k\in\cK}\!P_\gamma(k)\Var_k[(\widehat F_k)_1]$,
where the inequality arises from setting $M_k=[P_\gamma(k)/P_\cK]M$; a more optimal allocation of $M_k$s can only reduce the variance. Thus, 
\begin{align}
&\Var[(\widehat F_\textrm{str})_M]\leq \frac{P_\cK^2}{M}\Exp_\cK\{\Var_k[(\widehat F_k)_1]\}\\
&\qquad\leq \frac{P_\cK^2}{M}{\left[\Exp_{\cK}\{\Var_k[(\widehat F_k)_1]\}+\Var_{\cK}\{\Exp_k[(\widehat F_k)_1]\}\right]}=\Var[\widehat F_M],\nonumber
\end{align}
where the subscript $\cK$ on $\Exp_\cK$ and $\Var_\cK$ indicates computation over $\{P_\gamma(k)/P_\cK\}$ (conditioned on being in $\cK$), while $\Var[\widehat F_M]$ is computed over the unstratified distribution (still conditioned on $\cK$) $\{(P_\gamma(k)/P_\cK)p(\veci_\veca|k)\}$. From this, we see that stratified sampling gives significantly smaller variance whenever $\Var_{\cK}\{\Exp_k[(\widehat F_k)_1]\}=\Var_{\cK}(F_k)$, the variance of $F_k$ over $k$, is large. 

Next, let us discuss the choice of conditional proxy distributions and the sample sizes $M_k$s. Now, $p(\veci_\veca|k)=p(\veci_\veca|\veca)p(\veca|k)$, with $|\veca|=k$. Since $p(\veca|k)=\binom{A}{k}^{-1}$, the freedom of choice is in $p(\veci_\veca|\veca)=\prod_{a\in\veca}p_{i_a}^{(a)}$, which we use to minimize $\Var_k[(\widehat F_k)_1]$. Here, we follow the same logic as before for standard importance sampling [see the paragraph following Eq.~\eqref{eq:w}], but with $\cF^{(a)}$ and $r_i$s in place of the earlier $\cE^{(a)}$ and $q_i$s. As before, we make the choice to minimize $\Exp_k[w^2]$ and set 
\begin{equation}
p_{i_a}^{(a)}=\frac{|r_{i_a}^{(a)}|}{\beta^{(a)}}, \quad \textrm{with~}\beta^{(a)}\equiv 1+2\nu^{(a)},
\end{equation}
where $\nu^{(a)}\equiv \sum_{i_a,r_i<0}|r_{i_a}^{(a)}|$ is the negativity for $\cF^{(a)}$. For locations $a=0$ and $A+1$, the earlier prescription of Ref.~\cite{bennink2017} for the proxy distributions suffices. These locations do not participate in the stratification, and any choice made here is just for these two fixed locations, of no consequence as the circuit size grows. 

For given $p(\veci_\veca|k)$, it is easy to show that setting $M_k=MP_\gamma(k)\sigma_k/\sum_{k'\in\cK}P_\gamma(k')\sigma_{k'},$ minimizes the variance of the stratified estimator, where $\sigma_k\equiv \Var_k(F_k)$. We generally do not know $\sigma_k$s, but they can be estimated by small-sized pilot runs. In practice, it suffices to continue sampling until we see the variance of $\widehat F_k$ to be small enough so that $\widehat F_\textrm{str}$ converges to within some acceptable error. Note that one could potentially further reduce the variance by minimizing $\Var[(\widehat F_\textrm{str})]$ over the $p(\veci_\veca|k)$s and $M_k$s simultaneously, but we find the separate minimization used here already sufficient for our examples.

\subsection{Self-normalized importance sampling}\label{app:SamplingSNIS}
As mentioned in the main text, we make one further modification to the estimator for $F_k$, for easier numerical implementation: We replace the standard estimator of Eq.~\eqref{eq:FkMk} by the self-normalized importance sampling estimator. Self-normalised importance sampling (see, for example, Ref.~\cite{owen2013}) is useful when the probability distributions are known only up to a constant; that unknown constant cancels out in the ratio of estimators in $\FSN$. In our current setting, we do not have such an unknown constant. However, the ratio cancels the $\beta$ factors, so that
\begin{equation}
(\FSN)_{M_k}=\frac{\sum_{m=1}^{M_k}\mathrm{sgn}\{r(\veci_\veca^m)\}f(\veci_\veca^m)}{\sum_{m=1}^{M_k}\mathrm{sgn}\{r(\veci_\veca^m)\}}
\end{equation}
is independent of $\beta$, and hence also of $\gamma$ and $\nu$; only the signs of the $r$s enter. This independence of $\beta(>1)$ gives better numerical stability, especially for large $k$---$\beta^k$ can be quite large even for small negativity $\nu$, and this avoids having to handle the full range between $\pm\beta^k$.

$\FSN$ is asymptotically unbiased: $\lim_{M_k\rightarrow\infty} \FSN=\Exp[\widehat{F_k}]/\Exp[\widehat{w_k}]=F_k$ since $\Exp[\widehat{w_k}]=1$. It is, however, biased for finite sample sizes. We can estimate its bias in the large-$M_k$ limit with the so-called ``delta method" (see Ref.~\cite{verHoef2012} for the origins of this method), which looks perturbatively at deviations from the asymptotic limit (see App.~\ref{app:bias}),
\begin{align}
\mathrm{Bias}[(\FSN)_{M_k}]&\equiv \Exp_k[(\FSN)_{M_k}-F_k]\\
&\simeq\frac{1}{M_k}{\left[F_k\Var_k(w)-\Exp_k(w^2f)+F_k\right]}.\nonumber
\end{align}
This bias can be estimated from the samples themselves, and we stop when the bias is within the accepted error for the value of $F_k$. Note also that the bias vanishes asymptotically as $1/M_k$, while the standard deviation (square root of the variance) goes as $1/\sqrt{M_k}$ (see App.~\ref{app:bias}), so the bias shrinks more quickly than does the error bar around our estimate, when $M_k$ is large. Note also that one can use a proxy distribution optimized for the self-normalized estimator; however, we find our earlier proxy chosen based on $\widehat F_k$ already sufficient for our examples.

\subsection{Rejection sampling for nearby noise strengths}\label{app:RejectionSampling}
Suppose we have samples, labeled by $i$, drawn from the reference distribution $\widetilde P(i)$. The target distribution is $P(i)$. Assuming $P(i)\leq c\widetilde P(i)~\forall i$ for some constant $c$, let the acceptance probability be $p_\mathrm{acc}(i)\equiv P(i)/c\widetilde P(i)$. Then, for every sample $i$ drawn from $\widetilde P(i)$, we accept it with probability $p_\mathrm{acc}(i)$; the accepted samples will be distributed according to $P(i)$. 

For our stratified sampling, the target distribution is the joint distribution formed from the $\{p_{i_a}^{(a)}(\gamma)\}$s (given $k$), for some noise strength $\gamma$. For each configuration $\veci_\veca\in\cC_k$, chosen according to $\{p_{i_a}^{(a)}(\gamma)\}$s, we perform a stabilizer circuit simulation to calculate $f$. Now, $f(\veci_\veca)$ does not depend on $\gamma$; $\gamma$ determines only the probability of choosing $\veci_\veca$ (given $k$). Each pair $\{\veci_\veca,f(\veci_\veca)\}$, regarded as a single sample point, is thus valid for any value of $\gamma$. If we have generated a pool of such pairs for some value of $\gamma$ using stabilizer simulation, we can apply rejection sampling to get samples for a different value of $\gamma$, without going through the stabilizer circuit simulation again.

As mentioned in the main text, for rejection sampling to do well, the acceptance probability has to be reasonably large (and easily computable, which it is here), so that we do not need to generate a significantly larger pool of samples from $\widetilde P(i)$ to get sufficient $P(i)$ samples. This happens when $P$ and $\widetilde P$ are close enough so that $c$ is not too large, and $P(i)/\widetilde P(i)$ values are not too small. Functionally, this translates into having similar $\gamma$ values between the reference and target.

\section{Bias and variance for $\FSN$}\label{app:bias}
We can derive the expressions for the large-sample bias and variance of the self-normalized importance sampling estimator $\FSN$ using the delta-method. Suppose we have two real random variables $V_1$ and $V_2$; we let $W\equiv g(V_1,V_2)$, some function of the two variables. Let $\mu_i\equiv \Exp(V_i)$, $\lambda_i\equiv V_i-\mu_i$, $\sigma_i^2\equiv \Var(V_i)=\Exp(\lambda_i^2)$, and $\sigma_{12}\equiv \textrm{co}\Var(V_1,V_2)=\Exp(\lambda_A\lambda_B)$ the covariance between the two variables. Note that $\Exp(\lambda_i)=0$. We use the shorthand $\mu$ to denote both $\mu_i$s.

We consider the situation where $\lambda_i$s are small, and Taylor expand $W$ about $\lambda_i=0$, or, equivalently, about $V_i= \mu_i$:
\begin{align}
W&=g(V_1,V_2)=g(\mu_1,\mu_2)+\sum_{i=1,2}{\left.\frac{\partial g}{\partial V_i}\right\vert}_{V_i=\mu_i}\lambda_i\\
&\qquad +\frac{1}{2}\sum_{i=1,2}{\left.\frac{\partial^2 g}{\partial V_i^2}\right\vert}_\mu\lambda_i^2+{\left.\frac{\partial^2g}{\partial V_1\partial V_2}\right\vert}_\mu\lambda_1\lambda_2+\ldots.\nonumber
\end{align}
In our setting, we consider $g(V_1,V_2)=V_1/V_2$, so that
\begin{equation}
W\simeq \frac{\mu_1}{\mu_2}+\frac{1}{\mu_2}{\left[\lambda_1-\frac{\mu_1}{\mu_2}\lambda_2\right]}+\frac{1}{\mu_2^2}{\left[\frac{\mu_1}{\mu_2}\lambda_2^2-\lambda_1\lambda_2\right]},
\end{equation}
and the expectation value can be approximated as
\begin{equation}
\Exp(W)\simeq \frac{\mu_1}{\mu_2}+\frac{1}{\mu_2^2}{\left[\frac{\mu_1}{\mu_2}\sigma_2^2-\sigma_{12}\right]}.
\end{equation}

There remains the choice of the proxy distributions $p(\veci_\veca|k)$s and the sample sizes $M_k$s. Now, in our case, $W=(\FSN)_{M_k}$ so that $V_1=(\widehat F_k)_{M_k}$ and $V_2=(\widehat w_k)_{M_k}$, and we have $\mu_1=F_k$, $\mu_2=1$, $\sigma_1^2=\frac{1}{M_k}\Var_k[w(\veci_\veca)f(\veci_\veca)]$, $\sigma_2^2=\frac{1}{M_k}\Var_k[w(\veci_\veca)]$, and $\sigma_{12}=\frac{1}{M_k}\textrm{co}\Var_k[w(\veci_\veca)f(\veci_\veca),w(\veci)]=\frac{1}{M_k}{\left[\Exp_k(w^2 f)-\mu_1\mu_2\right]}$. 
Then, the expectation value is 
\begin{equation}
\Exp(W)\simeq F_k+\frac{1}{M_k}{\left[F_k\Var_k(w)-\Exp_k(w^2f)+F_k\right]},
\end{equation}
so that the bias for the self-normalized importance sampling estimator $(\FSN)_{M_k}$ is
\begin{equation}
\textrm{Bias}_k(\FSN)\equiv \Exp(W)-F_k
\simeq\frac{1}{M_k}\Exp_k[w^2(F_k-f)].
\end{equation}
A similar analysis gives its variance,
\begin{align}
\label{app:VarFSN}\Var_k(\FSN)\equiv \Var(W)&\simeq \frac{1}{\mu_2^2}\sigma_1^2+\frac{\mu_1^2}{\mu_2^4}\sigma_2^2-2\frac{\mu_1}{\mu_2^3}\sigma_{12}\\
&=\frac{1}{M_k}\Exp_k[w^2(F_k-f)^2],\nonumber
\end{align}
noting that $\Exp_k(wf)=\sum_{\veci_\veca\in\cC_k}p(\veci_\veca|k)w(\veci_\veca)f(\veci_\veca)=\sum_{\veci_\veca\in\cC_k}q(\veci_\veca|k)f(\veci_\veca)=F_k$.


\section{Unitary versus nonunitary channels}\label{app:UnitVsNonunit}

\begin{figure}[!h]
\includegraphics[width=\columnwidth]{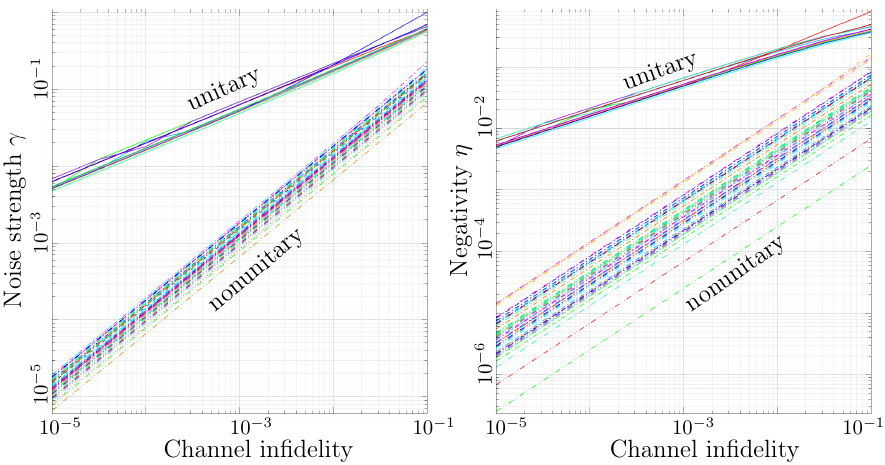}
\caption{\label{fig:UnitVsNonunit} Noise strength $\gamma$ and negativity $\eta$ as a function of the channel infidelity, for randomly chosen unitary and nonunitary channels. The solid lines are for unitary channels, while the dashed or dotted lines are for nonunitary channels.}
\end{figure}

In our simulations, we consistently found it more expensive to simulate unitary or coherent noise channels, compared to nonunitary channels. This can be attributed, at least in part, to the fact that the noise strength $\gamma$ and the negativity $\eta$ are observed to decrease more slowly for unitary channels as the channel infidelity falls, as shown in Fig.~\ref{fig:UnitVsNonunit}.

\section{Simulation details}\label{app:SimDetails}
Here, we provide further details on the sampling simulations used in our three examples.

\subsection{Computing $P_\gamma(k)$}\label{app:pk}
The stratified estimator requires the value of $P_\gamma(k)$, the probability of having exactly $k$ faulty locations, for $k\in\cK$. For our simple case of homogeneous noise strength, $P_\gamma(k)=\binom{A}{k} p^{\,k}(1-p)^{A-k}$, and this is available in standard numerical libraries. For the general inhomogeneous situation, with location $a$ having noise strength $\gamma^{(a)}$, the random variable is $K=\sum_{a=1}^A X_a$ with independent $X_a\sim\mathrm{Bernoulli}(\gamma^{(a)})$. The distribution of $K$ is then \emph{Poisson--binomial}:
\begin{equation}\label{eq:poisson_binom_def}
P_\gamma(K=k)
=\sum_{\substack{S\subseteq\{1,\dots,A\}\\ |S|=k}}
\prod_{a\in S} \gamma^{(a)}\prod_{b\notin S} (1-\gamma^{(b)}),
\end{equation}
with $k=0,1,\ldots,A$. This has no general closed form but can be computed efficiently by dynamic programming or FFT-based convolution.

\subsection{Selecting the faulty locations}\label{app:fault_location_selection}
Each stratum $\cC_k$ contains all circuit configurations with exactly $k$ faulty locations.
To generate samples $\veci_{\veca}\in\cC_k$, we must first select the subset of locations
$\veca=\{a_1,a_2,\ldots,a_k\}\subseteq\{1,\ldots,A\}$ on which faults occur, before sampling the fault types $\{i_{a_j}\}$ at those locations. For homogeneous noise strengths, this can be done simply by choosing the faulty locations uniformly randomly among the $\binom{A}{k}$ subsets of locations. For example, one can generate a random permutation of the $A$ circuit locations, and take the first $k$ elements of the permutation of $\veca$. This has $O(A)$ time cost and is unbiased since each subset of size $k$ is equally likely. 

In the case of inhomogeneous noise strengths, the distribution over faulty subsets $\veca$ is no longer uniform. Instead, each subset carries weights determined by $\gamma^{(a)}$s, with locations of larger $\gamma^{(a)}$ more likely to be faulty. In this case, sampling $\veca$ with $|\veca|=k$ amounts to drawing $k$ distinct indices from $\{1,2,\ldots A\}$ without replacement, weighted proportionally to their respective noise strengths $\gamma^{(a)}$. This can be done efficiently via the Efraimidis-Spirakis algorithm \cite{efraimidis2006}, with a complexity of $O(A\log k)$ and memory $O(A)$.

\subsection{Sample allocation across strata and error bars}\label{app:SampleAllocation}

For a given $\gamma$ value, stratum estimators $\FSN$ are estimated in decreasing order of $P_\gamma(k)$, stopping once the remaining unvisited $k$ values contribute negligibly, within the desired precision, to $\widehat F_M$. In our examples, we also observe $\FSN$ to vary smoothly with $k$. This allows us to also then sample only a subset of the $k$ values in the relevant set $\cK$, and interpolate between those sampled $k$ values. This is explained in detail in the surface-code example (App.~\ref{app:Interpolate}), where we make use of this interpolation to substantially reduce the simulation costs.

The total variance of our estimator $\widehat F_M$, with total samples $M=\sum_kM_k$, is given by 
\begin{equation}
\Var[\widetilde F_M]=\sum_{k=0}^A P_\gamma(k)^2\Var_k[\FSN], 
\end{equation}
with $\Var_k[\FSN]$ from Eq.~\eqref{app:VarFSN}; $M=\sum_kM_k$ is the total sample size across strata, assumed to be large. For fixed $M$, the total variance is minimized by allocating per-stratum sample size as
\begin{equation}
M_k\propto P_\gamma(k)\sqrt{\Var_k[\FSN]}.
\end{equation}
Short pilot runs are used to estimate the $\Var_k[\FSN]$s and guide adaptive allocation, updating $M_k$s until estimated contributions to $\Var[\widetilde F_M]$ are approximately balanced across strata. 

In our examples, the error bars shown in the plots are the estimated standard deviation $\sqrt{\Var[\widetilde F_M]}$ values, with $M$ large enough to achieve 10\% confidence interval of the estimated mean, with a 99\% confidence level.

\section{Further details for the surface-code example}
\subsection{Worst-performing channel for surface code}\label{app:WorstE}

As seen in Fig.~\ref{fig:SurfCodeAll}, the channel with the largest logical infidelity we found (for the $d=3$ situation), among the channels we tested, was a one-parameter family of nonunitary channels, with three Kraus operators, $E_1\equiv\sqrt{\varepsilon}E_1'$, $E_2\equiv \sqrt{\varepsilon}E_2'$, and $E_0\equiv \sqrt{\id-E_1^\dagger E_1-E_2^\dagger E_2}$, where $E_1'$ and $E_2'$ are the following randomly generated matrices:
\begin{align}
E_1' &=
\begin{pmatrix}
7.513196\times 10^{-2} + 2.284803\times 10^{-1}\,\mathrm{i} \\
-3.740817\times 10^{-2} - 5.503615\times 10^{-1}\,\mathrm{i} \\
8.559614\times 10^{-2} - 3.434704\times 10^{-2}\,\mathrm{i} \\
-4.949497\times 10^{-1} - 4.828740\times 10^{-1}\,\mathrm{i}
\end{pmatrix},\nonumber\\
E_2' &=
\begin{pmatrix}
8.930400\times 10^{-4} + 2.591419\times 10^{-1}\,\mathrm{i} \\
-1.370156\times 10^{-1} + 4.818675\times 10^{-1}\,\mathrm{i} \\
7.107620\times 10^{-2} + 9.455447\times 10^{-2}\,\mathrm{i} \\
-1.669140\times 10^{-1} + 4.662648\times 10^{-2}\,\mathrm{i}
\end{pmatrix}.
\end{align}
$\varepsilon$ is the noise strength parameter.

\subsection{Interpolating the fidelity curves}\label{app:Interpolate}
The logical fidelity is expected to decrease from 1 when we move away from the region with $k$ smaller than the number of correctable errors promised by the code. We assume the decrease to be monotonic, starting with an $F=1\equiv F_\textrm{small}$ plateau for small $k$, and an $F=0.5\equiv F_\textrm{large}$ (completely mixed output logical state) plateau at large $k$ \footnote{$F$ can, in principle, be taken to saturate at some other value at large $k$, if there is additional information. For large $k$, the variance of $F_k$ can be very large for large $d$, but a rough estimate suffices since those points contribute only a tiny amount because of the small value of $P_\gamma(k)$ there, and a small shift changes the interpolation only slightly. For the Steane-code case, we could directly verify that $F_A=0.5$ for the worst-case state (note that Fig.~\ref{fig:SteaneCode} shows a large-$k$ value of $F_k=0.6$ because that is the average, not worst-case, over the 6 Pauli eigenstates).}. Because of the smoothness of $F_k$, we need not sample for every relevant $k$; instead, we use an adaptive strategy that first tests whether $F_k$ is flat, and if not, concentrates the sampling around the transition. 

Specifically, we begin by sampling at $k=t+1$ and at $k=A$, with $A$ as the total number of fault
locations. We draw independent batches to obtain $\widehat F_{t+1}$ and $\widehat F_A$, together with their sampling uncertainties. If both estimates have low enough variance for our target overall precision, and are consistent with a common constant value $c_0$, then, under the monotonicity assumption, we regard $\widehat F_k\approx c_0$ for all $k\in[t+1,A]$ and no further sampling is required. If instead $\widehat F_{t+1}$ and $\widehat F_A$ are sufficiently well determined but differ, we know that a transition lies between $t+1$ and $A$. We approach the transition from both sides by doubling and halving: Starting from $k=t+1$, we set $k_{j+1}=2k_j$ while $\widehat{F}_{k_j}$ remains close to $\widehat{F}_{t+1}$, and starting from $k=A$, we set $k_{j-1}=k_j/2$ while
$\widehat{F}_{k_j}$ remains close to $\widehat{F}_A$. Once a deviation from a plateau is observed on either side, we identify the last fault count still consistent with the plateau, $k_{\mathrm{pl}}$, and the first that deviates, $k_{\mathrm{dev}}$, and perform a bisection search on the
interval $[k_{\mathrm{pl}},k_{\mathrm{dev}}]$ to localize more precisely the
value of $k$ where $\widehat F_k$ deviates from the plateau, and to improve the estimate of the curvature of the transition. 

This adaptive sampling procedure allows us to identify the small set of $k$ values where $F_k$ begins to transition between the two plateau values. However, in many scenarios, the fidelities at large $k$ can have large variance; in this case, we can apply only the doubling procedure, approaching the transition from the left side, with the assumed plateau of $\widehat F=0.5$ on the right. This, together with the assumption (observed empirically) that the transition is symmetric around the midpoint of the transition, allows us to get at the larger $k$ values. This works well enough for the level of precision needed in our examples. 
 
\begin{figure}
\includegraphics[trim=15mm 35mm 450 30mm,clip,width=\columnwidth]{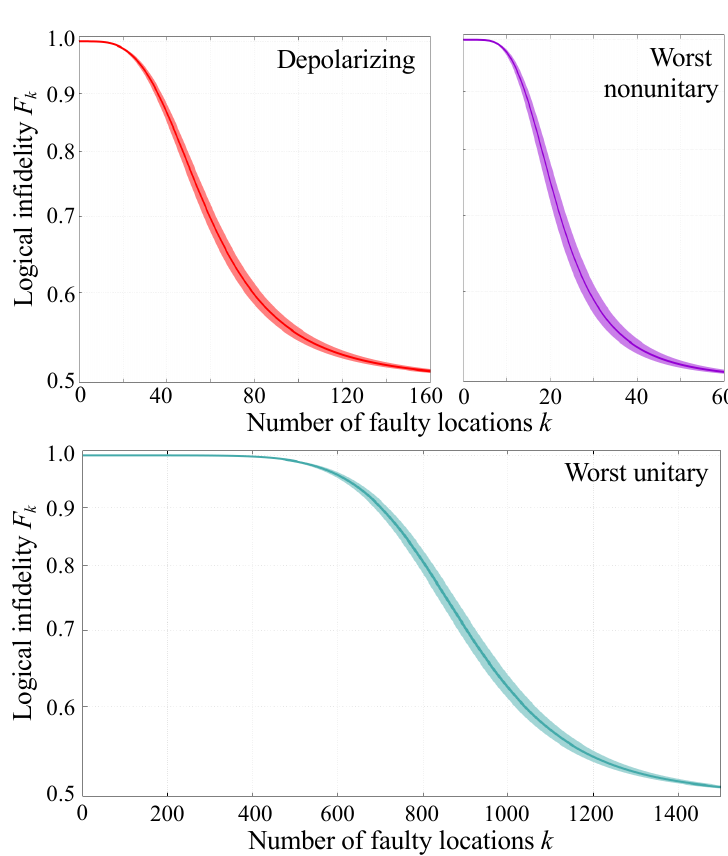}
\caption{\label{app:SurfCodeData} The estimated logical infidelity $F_k$ versus number of faulty locations $k$, for three of the channels in the surface-code example, for $d=7$. The shaded regions are the error bars, which encapsulate the uncertainty from the sampling as well as the interpolation. The sampled points are given in Tab.~\ref{tab:SurfCodeData}. More details on the interpolation are provided in App.~\ref{app:Interpolate}.}
\end{figure}

The remaining task is to interpolate monotonically between $F=1$ for small $k$ and $F=0.5$ for large $k$ in a way consistent with the sampled points and
their uncertainties. We do this by fitting a monotonic S-curve to the data: We use a standard four-parameter logistic function of the form
\begin{equation}
F_k=F_\textrm{large} + \frac{F_\textrm{small}-F_\textrm{large}}{1+{\left(\frac{k-B}{C}\right)}^D},
\end{equation}
where $F_\textrm{small}=1$ and $F_\textrm{large}=0.5$, and the shape parameters \(B,C,D\) are obtained by weighted nonlinear least squares, with weights proportional to
$1/\mathrm{Var}[\widehat{F}_k]$ so that more precise points dominate the fit.
The resulting parameter covariance matrix induces uncertainty bands for $F_k$ over the entire relevant $k$ range: At each $k$, the error
bar quantifies how far the monotone interpolation between $F_\textrm{small}$ and $F_\textrm{large}$ can be shifted while still remaining compatible with the
estimated values $\widehat{F}_k$ and their uncertainties. These are the error bars shown in Fig.~\ref{app:SurfCodeData}.

\begin{table}
\begin{tabular}{cc|cc|cc|cc|cc|cc}
\multicolumn{2}{c|}{Depol}&\multicolumn{2}{c|}{AD}&\multicolumn{2}{c|}{$NU_\textrm{best}$}&\multicolumn{2}{c|}{$NU_\textrm{worst}$}&\multicolumn{2}{c|}{$U_\textrm{best}$}&\multicolumn{2}{c}{$U_\textrm{worst}$}\\
\hline\hline
$k$&$\frac{M_k}{1000}$&$k$&$\frac{M_k}{1000}$&$k$&$\frac{M_k}{1000}$&$k$&$\frac{M_k}{1000}$&$k$&$\frac{M_k}{1000}$&$k$&$\frac{M_k}{1000}$\\
\hline
4&1&4&1&4&1&4&1&4&1&4&1\\
8&1&8&1&8&1&5&1&8&1&8&1\\
9&1&14&1&16&1&6&1&16&1&16&1\\
10&1&15&1&32&1&8&2.5&32&1&32&1\\
12&1&16&1&34&1&16&3&64&1&64&1\\
16&1&32&2&35&1&&&128&1&128&1\\
32&1&64&3&36&1&&&256&1&256&1\\
&&&&40&1&&&512&1&320&1\\
&&&&48&1&&&1024&1&352&1\\
&&&&64&2&&&1056&1&368&1\\
&&&&128&3&&&1072&1&369&1\\
&&&&&&&&1080&1&370&2\\
&&&&&&&&1084&1&372&2\\
&&&&&&&&1085&2&376&2\\
&&&&&&&&1086&2&384&4\\
&&&&&&&&1088&2&512&3.5\\
&&&&&&&&1152&4\\
&&&&&&&&1280&7\\
&&&&&&&&1536&6\\
&&&&&&&&2048&8\\
\hline\hline
$\frac{M}{1000}$&7&&10&&14&&8.5&&44&&24.5

\end{tabular}
\caption{\label{tab:SurfCodeData} Parameter choices for the reference sample pools for the $d=7$ surface-code example, at $\epsilon_\textrm{ref}=10^{-5}$; numbers at other $\epsilon_\textrm{ref}$ values are very similar. Shown in the table are the number of faults $k$ and the corresponding sample size $M_k$, used to generate the reference sample pools for the six channels in Fig.~\ref{fig:SurfCodeAll}: depolarizing (Depol), amplitude-damping (AD), best nonunitary ($NU_\textrm{best}$), worst nonunitary ($NU_\textrm{worst}$), best unitary ($U_\textrm{best}$), and worst unitary ($U_\textrm{worst}$) channel). The bottom line $M$ gives the total number of samples $M$ for each channel.
}
\end{table}

\begin{table}[h!]
\centering
\begin{tabular}{c||c|c|c|c|c|c}
$d=$ & 3 & 5 & 7 & 9 & 11 & 15 \\
\hline\hline
Depol               & 1(4) & 1(5) & 1(7) & 1(9) & 1(10) & 1(13)  \\
AD                  & 4 &2 & 2 & 2 & 3 & 4 \\
$NU_\textrm{best}$  & 5 &4 & 2 & 3 & 4 & 3 \\
$NU_\textrm{worst}$ & 6 & 3 & 2 & 2 & 2 & 2 \\
$U_\textrm{best}$   & 8 & 7 & 7 & 6 & 7 & 6 \\
$U_\textrm{worst}$  & 5 & 5 & 4 & 4 & 5 & 5 \\
\end{tabular}
\caption{\label{tab:Mvalues}Total sample sizes $M$, for a given $\epsilon_\textrm{ref}$, for different distances $d$ and different noise channels, for the surface-code example. The entries are given in multiples, rounded up to the nearest integer, of the $M$ for the depolarizing noise (top row); the numbers in parentheses in the top row are the actual $M/1000$ values for depolarizing noise. These numbers are representative for all $\epsilon_\textrm{ref}$ values.}
\end{table}

\subsection{Simulation costs}\label{app:SurfCodeCosts}

Here, we provide the data for estimating the resource costs of simulating the surface-code examples.
As in the Steane-code example, for a given target precision, the set of relevant $k$ values---those that contribute substantially to the fidelity estimate---depends on the product of the probability $P_\gamma(k)$ and the corresponding $F_k$ value. In addition, as explained in the main text, we also only need to sample a small number of those relevant values, sufficiently for us to do a good interpolation. Table \ref{tab:SurfCodeData} gives the actual $k$ values sampled for the $d=7$ example, and Fig.~\ref{app:SurfCodeData} shows the fidelity curves for the different channels. Table \ref{tab:Mvalues} gives the total sample sizes $M$ for all code distances in our example, to provide a comparison of relative simulation costs across the different noise channels.

\end{document}